\documentclass[a4paper,11pt]{article}
\usepackage{jcappub} 
\usepackage{lineno}
\linenumbers

\arxivnumber{1234.56789} 
\title{\boldmath A generalized solution for anisotropic compact star model in $F(\mathcal{Q})$ gravity}

\author[a]{Sat Paul,\note{Corresponding author.}}
\author[a]{Jitendra Kumar,}
\author[b]{Sunil Kumar Maurya,\note{Also at Some University.}}
\author[a]{Sourav Choudhary,}
\author[a]{Sweeti Kiroriwal}

\affiliation[a]{Department of Mathematics, Central University of Haryana, Jant-Pali, Mahendergarh 123029, Haryana, India}
\affiliation[b]{Department of Mathematical and Physical Sciences,
College of Arts and Sciences,\\ University of Nizwa, P.O. Box 33, Nizwa 616, Sultanate of Oman}
\emailAdd{satpaul232039cuh@gmail.com}
\emailAdd{jitendark@gmail.com}
\emailAdd{sunil@unizwa.edu.om}
\emailAdd{chaudharysourav192@gmail.com}
\emailAdd{sweeti222020@cuh.ac.in}

\abstract{In this work, we investigate an anisotropic compact star's physical properties and stability in $F(\mathcal{Q})$ gravity. The study focuses on the significance of $F(\mathcal{Q})$ gravity on the structure and stability of compact star, considering non-perfect fluid. Buchdahl ansatz along with transformation used to solve the Einstein field equations. We investigate the physical parameters of the 4U1820-30 compact star using a static spherical metric in the interior region and a Schwarzschild (anti) de-sitter metric in the exterior region. We investigate the behaviour of energy density($\rho$), radial pressure($p_{r}$), tangential pressure($p_{t}$), anisotropy($\Delta$), metric potentials, energy state parameter, and energy requirements in the interior of the proposed stellar object. The equilibrium state of this star is analysed using the Tolman-Oppenheimer-Volkoff(TOV) equation and their stability is determined using the regularity condition, causality condition, the adiabatic index($\Gamma$) method, and Herrera cracking method.}

\begin{document}
\nolinenumbers
\maketitle
\flushbottom

\section{Introduction}\label{sec1}
\label{sec:intro}
Most galaxies have an erratic distribution of stars formed in clouds of gas and dust with an uneven matter distribution. All active stars ultimately come to a stage in their evolution where the constant gravitational pull is no longer resisted by the radiation pressure originating from the nuclear fusions occurring within it. According to~\cite{Sagert:2005fw}, stellar death occurs when a star collapses under its weight. This process produces an extremely dense and compact stellar remnant, known as a compact star. Particularly, compact stars are the last phases of the evolution of ordinary stars, including white dwarfs, neutron stars, black holes, and quark stars. The evolution of compact stars is related to the phase separations that occurred in the early universe. Compact stars are identified by the stop of nuclear processes in their interiors, in addition to their extreme density. This is the reason they are unable to resist the force of gravity. Gravity in white dwarfs and neutron stars is resisted by the pressure of degenerate gases. In black holes, gravity dominates over other forces, compressing star material to infinite density~\cite{karttunenfundamental}. The outer layers are expelled and when a star like the Sun runs out of its nuclear fuel, the core collapses to produce a compact and white star known as the white dwarf. When inert iron cores of stars with masses equal to or greater than ten times that of the Sun collapse, they form an extremely dense neutron star or even a black hole.
Physicists began utilising stars as ``Physics Laboratories" in the mid-1920s to study how matter transforms under tremendous pressures and densities a phenomenon that is not possible to observe on Earth. In 1917 Bonolis provided details on how the concept of compact stars evolved. Oppenheimer initiated the initial conversation regarding the potential for gravitational collapse. In 1915, white dwarfs offered a novel means of testing Einstein's theory of general relativity (GR)~\cite{o1996general}, but the neutron star hypothesis, when combined with observations of supernovae, produced an even more extensive test of the theory. Eddington went into great depth about white dwarfs in 1924. Additionally, Friedrich divided the ultra-relativistic electron gas in the very dense stars into two groups: non-relativistic and ultra-relativistic. He made an accurate estimate of the mass of a stable star, which is limited to a specific maximum M($\geqslant M \odot$) and exists in a relativistic degenerate form. Some basic equations have been established to regulate the configuration of a star in radiative equilibrium. In 1926, Fowler~\cite{fowler1926general} explored the issue of degenerate dense matter in white dwarf stars. Oppenheimer used a variety of theoretical frameworks to contribute to the relativistic gravitational collapse of a neutron star toward the end of the 1930s. These studies initiated a new line of inquiry into the relationship between GR and compact objects. In 1926, Fowler discovered that the connection between white dwarfs' density, energy, and temperature could be determined by studying them as a gas of non-relativistic, non-interacting nuclei that follow Fermi-Dirac statistics~\cite{fowler1926general,Fowler:1926zz}. In 1929 stoner utilised the Fermi gas model to determine the mass, radius, and density of white dwarf stars, assuming they were homogenous spheres of electrons. Wilhelm Anderson modified this model using relativity and proposed a theory in December 1928 that said ``the mass of a star must have a maximum value" \cite{anderson1929limiting}. Stoner deduced the internal energy density equation of state for Fermi gases in 1930 and he was lastly able to approach the mass-radius equation in a completely relativistic manner as a result, and he came to the conclusion that its limiting mass is around $2.19\times10^{30}$ kg~\cite{stoner1930lxxxvii}. Chandrasekhar published many papers \cite{Chandrasekhar:1931ftj,chandrasekhar1974black,chandrasekhar1934physical,Chandrasekhar:1934in,Chandrasekhar:1935zz,chandrasekhar1935stellar}  between 1931 and 1935. In these publications, he solved both the hydrostatic and nonrelativistic Fermi gas equations of state. Additionally, he contributed to the relativistic Fermi gas scenario, which resulted in the value of the first predetermined boundary. The Chandrasekhar limit is the name given to this new limit. The prevailing consensus regarding the Chandrasekhar 1.4 solar masses is the upper limit~\cite{ Mazzali:2007et}. In 1938 Zwicky discovered the idea of neutron star ~\cite{Baade:1934onh,baade1934remarks}  and made a fresh effort using GR in this direction. He established a more strong link between GR and the idea that the mass of a star of a given density cannot exceed a particular crucial value known as the Schwarzschild limit~\cite{zwicky1938neutron}. He presented his ideas and critiqued the Schwarzschild solution~\cite{schwarzschild2003gravitational}.In his conclusion, he states that any star that reaches the Schwarzschild limiting configuration must be viewed as an object that effectively prevents physical connection with the rest of the universe. As a result, it is difficult to see any physical states in stellar bodies that have crossed the Schwarzschild limit ~\cite{zwicky1938neutron}. 
 By accounting for the effects of nuclear forces~\cite{Oppenheimer:1938zz}, we were able to arrive at a fair estimate of the lowest mass for which such an interior would be stable (about $0.1M \odot$).
In 1939, Oppenheimer and Volkoff discovered that if the total mass of neutrons exceeds around $0.7M \odot$, the general relativistic field equations cannot offer a static solution for a spherical distribution of cold neutrons.  Oppenheimer~\cite{Oppenheimer:1939ne} proved that a star under these conditions would collapse due to its gravitational field. The TOV limit is the value that sets a limit on the mass of a cool, non-rotating neutron star and this limit was comparable to the white dwarf star mass at the Chandrasekhar limit. Given the high nuclear repulsion forces that exist between neutrons, current research suggests that this limit may be anywhere from 1.5 and $3M \odot$ ~\cite{bombaci1996maximum}. Nonetheless, it was observed that rotation could raise the maximum mass ~\cite{friedman1987maximum, Cook:1993qj, eriguchi1994structure, Paschalidis:2016vmz}. Scientists Snyder and Oppenheimer had studied gravitational collapse. The whole ramifications of Einstein's theory of gravitation are included in this work ~\cite{Oppenheimer:1939ue}. A small group of scientists had begun working on the issue of what would happen to a compact star interior composed only of neutrons by 1939. Due to these contributions, it became clear that only Einstein's theory of gravity could account for such extraordinarily dense materials, while on the one hand, nuclear matter and particle physics were becoming increasingly important in characterizing matter at such high densities. The science of relativistic astrophysics becomes understandable.
Though revolutionary, GR has faced difficulties in understanding the Universe's late-time acceleration, the existence of dark matter, and the quantization of gravitational interaction. The existence of dark matter and the quantization of gravitational interaction are two of the many unresolved problems that must be addressed to describe the Universe's late-time acceleration. There are disadvantages to using the affine connection in GR, This discrepancy in the study can be explained by selecting different affine connections with various gravity descriptions ~\cite{ellis2012relativistic, BeltranJimenez:2019esp, Harada:2020ikm}. This implies alternate theories of gravity, such as the modified theory of gravity. Some modified gravity theories, like $F(\mathcal{R})$ gravity, $F(\mathcal{R}, T)$ gravity, $F(\mathcal{Q})$  gravity, teleparallel gravity, and uni modular gravity so on. Modified metrics can reveal different circumstances, which can be explored further in future research ~\cite{Wang:2021zaz, Errehymy:2022hxt, Kaur:2023gca, Astashenok:2020qds, Astashenok:2021peo, Maurya:2020rny, Maurya:2019hds, Kaur:2022zmi}.
One of the more plausible ideas to address the shortcomings of GR is the  $F(\mathcal{R} )$ theory. Scalar curvature in the Lagrangian can be substituted to provide $F(\mathcal{R})$  gravity. The usefulness of this idea in tackling dark energy is demonstrated
in various research paper~\cite{de2010f,Sotiriou:2008rp,Capozziello:2011et}. Similar to how $F(T)$ gravity can be created, the affine connection in GR allows the requirement that the curvature $(\mathcal{R})$  can be nonzero while allowing the values of nonmetricity $(\mathcal{Q})$  and torsion  $(T)$  to vanish. It is possible to build a wide range of theories and geometries with $\mathcal{Q}$, $\mathcal{R}$, and $T$ in many mergers, such as zero or non-zero. As was previously mentioned, when we choose zero curvature and nonmetricity in the case of $F({T})$ gravity, we have a symmetric teleparallel formation(STGR). Various studies in the field of (STGR) have been conducted and are included in these papers ~\cite{Nester:1998mp,Adak:2005cd, Adak:2008gd,Mol:2014ooa,Jarv:2018bgs, BeltranJimenez:2017tkd,BeltranJimenez:2019tme, Gakis:2019rdd,Li:2022vtn,Li:2021mdp,LiMingZhe:2021vtu,Agrawal:2022vdg,Capozziello:2021pcg}. An arbitrary function of $(\mathcal{Q})$ is contained in the gravitational Lagrangian to create $F(\mathcal{Q})$ gravity. In $F(\mathcal{Q})$ gravity's field equations are second-order, which is a significant advantage over the $F(\mathcal{R})$ theory's fourth-order differential equations.  $F(\mathcal{Q})$ gravity was the primary focus of this investigation. Modified gravity, which increases the inertial gravitational interaction, allows for the easy creation of classical GR without taking into account the affine space-time structure. $F(\mathcal{Q})$ gravity is unusual in that it can account for the Universe's late-time acceleration without changing the scalar fields.~\cite{Akarsu:2010zm,BeltranJimenez:2019tme,Lazkoz:2019sjl}. A wide range of disciplines, including wormhole geometry~\cite{Harko:2018gxr,Xu:2019sbp,Xu:2020yeg,DAmbrosio:2020nev,Maurya:2022hmw,Lohakare:2023ugg,Capozziello:2022tvv,Banerjee:2021mqk} and dynamical analysis~\cite{Lu:2019hra,Mandal:2020buf}, which includes the topic of cosmography, have been used to establish the existance of the $F(\mathcal{Q})$ gravity model. There have also been several completely distinct methods to the modified theory of gravity. For example, in ~\cite{Lazkoz:2019sjl}, the background analysis and observations were carried out by switching to the $F({z})$ theory of gravity after the investigation had begun using the $F(\mathcal{Q})$ theory. In the paper, ~\cite{Bajardi:2020fxh} used a bouncing model to tackle the issue. The functional form $F(\mathcal{Q})= Q + Q^{n}$ has been used in research on holographic dark energy ~\cite{Shekh:2021ule}. This study demonstrated that there is no Schwarzchild analogue solution for a static spherically symmetric solution using the non-vanishing $F(\mathcal{Q})$ function~\cite{BeltranJimenez:2019tme}.
A coincident gauge, like the zero value of the torsion restriction and non-metricity, can be utilised consistently ~\cite{BeltranJimenez:2017tkd}.
 This approach is distinguished by the fact that the affine connection vanishes. Using the coincident gauge ~\cite {Harko:2018gxr, Lu:2019hra, Frusciante:2021sio} and treating the metric as the only basic variable in modified (STGR) theories like the $F(\mathcal{Q})$  theory might lead to changes in metric development across different coordinate systems. Anisotropy affects the macroscopic features of a star, including mass, radius, tidal deformability, and nonradial oscillation ~\cite{Roupas:2020jyv, Deb:2021ftm, Estevez-Delgado:2018ydn, Pattersons:2021lci, Rizaldy:2019lwc, Rahmansyah:2020gar, Rahmansyah:2021gzt, Herrera:2007kz, Herrera:2013fja, Biswas:2019gkw, das2021role, Roupas:2020mvs}. Herrera and Santos ~\cite{Herrera:1997plx} thoroughly investigated anisotropic fluid spheres. Based on the preceding literature study, we will broaden our search for new physical solutions to Einstein's field equations for static and spherically symmetric anisotropic fluid distribution. The approach assumes an uncharged fluid sphere, highlighting the importance of anisotropy in modelling.
 We used metric potentials as generating functions and then calculated energy density and pressure. 
To compare with the observational data, we addressed the compact object 4U1820-30 in the structure of the model. \\The present work has been designed as follows:\\
Section~\ref{sec2}: In this section, we write the field equation of $F(\mathcal{Q})$ Gravity \\
Section~\ref{sec3}: In this section, we commenced our proposed model for an anisotropic star, obtained expression for energy density, radial pressure, tangential pressure anisotropy factor, etc.\\
Section~\ref{sec4}: In this section, we discuss regularity and reality condition in $F(\mathcal{Q})$ gravity for a well-behaved model.\\
Section~\ref{sec5}: In this section, we discuss boundary conditions which are very important to calculating the value of constants.\\
Section~\ref{sec6}: In this section, we analysed the model physically as well as graphically.\\
Section~\ref{sec7}: In this section, the research work is concluded.

\section{Field equations of $F(\mathcal{Q})$ gravity}\label{sec2}
The Levi-Civita affine connection is crucial in explaining and analysing general relativity. Its solution on the space-time manifold is compatible with the metric. Different manifolds and affine connections can be employed, and several theories for gravity can be offered ~\cite{Harada:2020ikm, BeltranJimenez:2019tme}. The Levi-Civita connection highlights the importance of nonmetricity $(\mathcal{Q})$ and torsion T, with curvature. Relaxing these constraints allows for non-Riemannian geometry theories with non-vanishing curvature, torsion, and nonmetricity. The $F(\mathcal{Q})$ gravity model was developed using these conditions, resulting in a gravitational Lagrangian with an arbitrary nonmetric function $(\mathcal{Q})$. The extension of $F(\mathcal{Q})$ gravity is extremely important in the expansion of the Universe. Jiménez et al.~\cite{BeltranJimenez:2019tme} was the first to coin the term symmetric teleparallel gravity, generally known as the $F(\mathcal{Q})$ gravity.Following 
 ~\cite{Zhao:2021zab} work on $F(\mathcal{Q})$ gravity, we investigate a generic metric space-time with independent metric tensors $g_{a b}$ and $\Gamma^{\sigma}_{a b}$. The equation below demonstrates the nonmetricity of the aforementioned connection:

 \begin{eqnarray} \label{eq2.1}
\mathcal{Q}_{\alpha ab}=\nabla_{\alpha}g_{ab} =\partial_{\alpha}g_{a b}-\Gamma^{\sigma}_{\alpha a}g_{\sigma b}-\Gamma^{\sigma}_{\alpha b}g_{a \sigma}.
\end{eqnarray}
These three components, in their generic form, are obtained by disassembling this affine relationship into independent components:

\begin{eqnarray}\label{eq2.2}
\Gamma^{\sigma}_{a b}=\{^{\sigma}{}_{a b}\}+\zeta^{\sigma}_{a b}+L^{\sigma}_{a b},
\end{eqnarray}
where $\{^{\sigma}{}_{a b}\}$ denotes the levi-Civita connection, $\zeta^{\sigma}_{a b}$ denote the Contortion tensor and $L^{\sigma}_{a b}$ denotes the disformation. $\{^{\sigma}{}_{a b}\}$ is solved by using the metric potential $g_{ab}$ :
\begin{eqnarray}\label{eq2.3}
\{^{\sigma}{}_{a b}\} \equiv \frac{1}{2}g^{\sigma \beta}(\partial_{a}g_{\beta b}+\partial_{b}g_{\beta b}-\partial_{\beta}g_{a b}),
\end{eqnarray}
 $\zeta^{\sigma}_{a b}$ is written as:
\begin{eqnarray}\label{eq2.4}
\zeta^{\sigma}_{a b} \equiv \frac{1}{2}T^{\sigma}_{\hspace{0.1cm}} {a b}+T_{(a\hspace{0.3cm}b)}^{\hspace{0.2cm}\sigma}
\end{eqnarray}
and $L^{\sigma}_{a b}$ is written  as:
\begin{eqnarray}\label{eq2.5}  
L^{\sigma}_{a b} \equiv \frac{1}{2}Q^{\sigma}_{a b}-Q_{(a\hspace{0.3cm}b)}^{\hspace{0.2cm}\sigma}.
\end{eqnarray}
The nonmetricity conjugate is defined as:

\begin{eqnarray}\label{eq2.6}
   F^{\alpha}_{a b} = \frac{-1}{4}Q^{\alpha}_{a b} + \frac{1}{2}Q^{\alpha}_{(a b)} + \frac{1}{4}({Q}^{\alpha} -\Tilde{Q}^{\alpha})g_{a b}-\frac{1}{4}\delta^{\alpha}_{(a b)}.
   \end{eqnarray}
The independent traces of the preceding equations are as follows:
\begin{eqnarray}\label{eq2.7}
Q_{\alpha} \equiv Q_{\alpha a}^{\hspace{0.2cm} a} \quad \Tilde{Q}_{\alpha} \equiv {Q}^{a}_{\hspace{0.1cm}\alpha\hspace{0.1cm} a},
\end{eqnarray}
now, the nonmetricity scalar is defined as:
\begin{eqnarray}\label{eq2.8}
Q = -\hspace{0.1cm} Q_{\alpha b a}F^{\alpha a b}.
\end{eqnarray}
The $F(\mathcal{Q})$ gravity is defined by the following action using Lagrange multiplets.:
\begin{eqnarray}\label{eq2.9}
S= \int\sqrt{-g}d^{4}x\Bigg[\frac{1}{2}F(Q)+\sigma^{\beta a b}_{\alpha}\mathcal{R}^{z}_{\beta a b} +\sigma^{a b}_{\alpha} T^{\alpha}_{a b}+\mathcal{L}_{n}\Bigg].
\end{eqnarray}
In Eq.~(\ref{eq2.9}), $g$ represents the determinant metric, $F(\mathcal{Q})$ denotes the nonmetricity of $\mathcal{Q}$, $\mathcal{L}_{n}$ represents the Lagrange density, and $\sigma_{\alpha}^{\beta a b}$ represents the Lagrange multipliers.
The action given in Eq.~(\ref{eq2.9}) has been used to derive the field equations concerning the metric given below 
\begin{eqnarray}\label{eq2.10}
T_{a b} =\frac{2}{\sqrt{-g}}\Delta_{\alpha}(\sqrt{-g}F_{Q}H^{\alpha}_{a b})\nonumber\\&&\hspace{-3.8cm} + \frac{1}{2}g_{a b}F+F_{Q}(H_{a \alpha \beta} Q_{b}^{\alpha \beta}-2 Q_{\alpha \beta a}H^{\alpha \beta}_{b}).
\end{eqnarray}

In equation Eq.~(\ref{eq2.10}),
the subscript $\mathcal{Q}$ denote the derivative of the function $F(\mathcal{Q})$ with respect to $\mathcal{Q}$,the energy momentum tensor is defined as :
\begin{eqnarray}\label{eq2.11}
T_{a b} \equiv \frac{2}{\sqrt{g}} \frac{\delta(\sqrt{-g})\mathcal{L}_{n}}{\delta g^{a b}}.
\end{eqnarray}
we get the following equations when we change equation Eq.~(\ref{eq2.9}) concerning affine connection:
\begin{eqnarray}\label{eq2.12}
\nabla_{\rho}\sigma_{\alpha}^{b a \rho}+\sigma_{\alpha}^{a b}=\sqrt{-g F_{Q}}H_{a b}^{\alpha}+J_{\alpha}^{a b}.
\end{eqnarray}
The hyper momentum tensor density is represented by the following equation:
\begin{eqnarray}\label{2.13}
J_{\alpha}^{a b}= \frac{-1}{2} \frac{\delta \mathcal{L}_{n}}{\delta T_{a b}^{\alpha}}.
\end{eqnarray}

To simplify equation Eq.~(\ref{eq2.12}) to the following equation, we use the anti-symmetry property of a and b in the lagrangian multiplier coefficients:
\begin{eqnarray}\label{eq2.14}
\nabla_{a}\nabla_{b}(\sqrt{-g}F_{Q}H^{ab}_{\hspace{0.3 cm}\alpha}+ J_{\alpha}^{\hspace{0.3 cm}a b})= 0.
\end{eqnarray}

if we consider$ \nabla_{a}\nabla_{b}\nabla_{\alpha}^{a b}=0$ we get :
\begin{eqnarray}\label{eq2.15}
\nabla_{a}\nabla_{b}(\sqrt{-g}F_{Q}H^{a b}_{\hspace{0.3 cm}\alpha})= 0.
\end{eqnarray}
 The affine connection without torsion or curvature has the following form:
\begin{eqnarray}\label{eq2.16}
\Gamma^{\alpha }_{a b}=\Bigg( \frac{\partial{u}^{u}}{\partial{\xi}^\sigma}\Bigg)\partial_{a}\partial{b}\xi^\sigma.
\end{eqnarray}
If we consider a specific coordinate selection that is well-known as a coincident gauge, for which $\Gamma^{\alpha}_{u v}= 0$. After that, the nonmetricity becomes:
\begin{eqnarray}\label{eq2.17}
Q_{\alpha a b}=\partial_{\alpha}g_{a b}.
\end{eqnarray}
The metric for a general static and spherically symmetric space is expressed as : 
\begin{eqnarray}\label{eq2.18}
ds^{2}=-e^{\upsilon(r)}dt^{2}+ e^{\sigma(r)}dr^{2}+r^{2}dw^{2}.
\end{eqnarray}

In the above equation $dw^{2}\equiv d\theta^{2}+sin^{2}{\theta} d{\phi}^{2}$. we get Eq.~(\ref{eq2.19}) by putting Eq.~(\ref{eq2.18}) in Eq.~(\ref{eq2.8}), then  $\mathcal{Q}$  is expressed in terms of r,
\begin{eqnarray}\label{eq2.19}
Q(r) = -\frac{2e^{-\sigma}}{r}\Bigg(\upsilon^{\prime}+\frac{1}{r}\Bigg),
\end{eqnarray}
where prime($\prime$) signifies the derivative with respect to r. An anisotropic fluid with spherically symmetric geometry has the following energy-momentum tensor:

\begin{eqnarray}\label{2.20}
T_{a b}=(\rho+p_{t})a_{a}a_{b}+p_{t}g_{a b}+(p_{r}-p_{t})b_{a}b_{b},
\end{eqnarray}
where, $a_{a}$ denotes four-velocity, and $b_{b}$ denotes the unitary space-like vector in the radial direction that fulfils $a^{a}a_{a}=-1,b^{a}b_{a}=1,a^{a}a_{a}=-1,a^{a}b_{b}=0$. $\rho(r)$ represents energy density, $p_{r}(r)$ represents the radial pressure in the direction of $b_{a}$, while the orthogonal represents the tangential pressure $(b_{a})$ is denoted by $p_{t}(r)$.The following are the independent components of the equations of motion Eq.~(\ref{eq2.10}) for anisotropic fluid  Eq.~(\ref{eq2.12}):
\begin{eqnarray}\label{eq2.21}
&&\hspace{-0.5cm} \rho=-\frac{F}{2}+F_{Q}\Bigg[Q+\frac{1}{r^{2}}+\frac{e^{-\sigma}}{r}(\upsilon^{\prime}+\sigma^{\prime})\Bigg],\\
&& \hspace{-0.6cm} p _{r} = \frac{F}{2}-F_{Q}\Bigg[Q+\frac{1}{r^{2}}\Bigg]\label{eq2.22},\\
&& \hspace{-0.6cm} p_{t} = \frac{F}{2}-F_{Q}\Bigg[\frac{Q}{2}-e^{-\sigma}\frac{Q}{2}-e^{-\sigma}\Bigg[\frac{\upsilon^{\prime\prime}}{2}+(\frac{\upsilon^{\prime}}{4}+\frac{1}{2r})(\upsilon^{\prime}-\sigma^{\prime})\Bigg]\Bigg]\label{eq2.23},\\
&& \hspace{-0.6cm} 0 = \frac{\cot\theta}{2}Q^{\prime}F_{Q Q}\label{eq2.24}.
\end{eqnarray}
Assuming a zero affine connection in the coordinate system, $F(\mathcal{Q})$ gravity Eq.~(\ref{eq2.9}) can be expressed as follows:
\begin{eqnarray}\label{eq2.25}
 \frac{\cot\theta}{2}Q^{\prime}F_{Q Q}=0,
\end{eqnarray}
Combining equation (\ref{eq2.10}) and motion equations yields $F_{Q Q} = 0$. This leads us to the conclusion that the function $F(\mathcal{Q})$ must be linear. Choosing nonlinear values for the function $F( \mathcal{Q})$ might result in conflicting equations of motion and solutions, especially for $F(Q=Q^{2})$. Using a nonlinear function of $F( \mathcal{Q})$ gravity will give inconsistent values for the equation of motion. To investigate and solve the nonlinear function of $F( \mathcal{Q})$ gravity, we require a more extended spherically symmetric metric for coincident gauges. Please visit ~\cite{Zhao:2021zab} for a more in-depth study of this topic. In this investigation, we used a linear function with $F_{Q Q} = 0$ and found that the spherically symmetric coordinate system (\ref{eq2.18}) corresponds to the affine connection $\Gamma^{a}_{ab}=0$.
To formulate a realistic compact stellar model, we need to discover a useful function of $F( \mathcal{Q})$. According to ~\cite{Wang:2021zaz}, matching the compact star solution with the Schwarzchild (Anti-) de sitter solution on the exterior boundary requires selecting a function that assumes $F_{Q Q} = 0$. Then, to determine the field equations, we utilize the function $F(\mathcal{Q})$, which is expressed as follows:
\begin{eqnarray}\label{eq2.26}
F_{Q Q}=0\Rightarrow F(\mathcal{Q})= \beta_{1}Q+\beta_{2}.
\end{eqnarray}
Using equations  Eq.~(\ref{eq2.25}) and  Eq.~(\ref{eq2.26}) with $\beta_{1}$ and $\beta_{2}$ as integration constants, the field equations for $F(\mathcal{Q})$ gravity can be stated as follows:
\begin{eqnarray}\label{eq2.27}
&& \hspace{-0.5cm} \rho= \frac{1}{2r^{2}}\Bigg[2\beta_{1}+2e^{-\sigma}\beta_{1}(r{\sigma}^{\prime}-1)-r^{2}\beta_{2}\Bigg],\\
&& \hspace{-0.5cm} p_{r}=\frac{1}{2r^{2}}\Bigg[-2\beta_{1}+2e^{-\sigma}\beta_{1}(r\upsilon^{\prime}+1)+r^{2}\beta_{2}\Bigg]\label{eq2.28},\\
&& \hspace{-0.5cm}
p_{t}=\frac{e^{-\sigma}}{4r}\Bigg[2e^{\sigma}r\beta_{2}+\beta_{1}(2+r\upsilon^{\prime})(\upsilon^{\prime}-\sigma^{\prime})+2r\beta_{1}\upsilon^{\prime\prime}\Bigg]\label{eq2.29}.
\end{eqnarray}
In Eqs.~(\ref{eq2.27}--\ref{eq2.29}) prime $(\prime)$ denotes the differentiation with respect to radial coordinate r and  $\rho$,  $p_{r}$, $p_{t}$, $\sigma$ and $\upsilon$ are unknowns which we are going to solve to get our desired result. For static and spherically symmetric fluid configurations, the general expression of the anisotropic factor, denoted as $\Delta=(p_{t}-p_{r})$, can be stated as follows: 
\begin{equation}\label{eq2.30}
\Delta(r)=\frac{\beta_{1}}{8\pi}\Bigg[e^{-\sigma}\Bigg(\frac{\upsilon^{\prime\prime}}{2}-\frac{\sigma^{\prime}\upsilon^{\prime}}{4}+\frac{(\upsilon^{\prime})^{2}}{4}-\frac{\upsilon^{\prime}+\sigma^{\prime}}{2r}-\frac{1}{2r^{2}}\Bigg)+\frac{1}{r^{2}}\Bigg].
\end{equation}
\section{Solution for new model}\label{sec3}
The five functions that describe fluid and its gravitation are $\sigma$, $\upsilon$, $\rho$, $p_{r}$ and $p_{t}$ and they are dependent on the radius r, as was mentioned in the above section. Since there are only three fundamental field equations, two of the characteristics above referred to as the generating function must be provided and the remaining three fluid characteristics must be discovered by the use of field equations. Various solutions can be found based on the provided pair of functions. This does not preclude the possibility of overlap in the solutions derived from several examples. The situation involving knowing $\rho$ and $\Delta$ is entirely generic. Since $\rho$ and $\Delta$ in this instance are generating functions, we have control over them and may select them to be regular, positive, and very simple. However, in this scenario, the following three functions are typically more complicated and not always physically realistic. Thus, it would be preferable if we choose $\sigma$ and $\upsilon$ beforehand. Even though this is the simplest scenario, we are forced to proceed by trial and error because we have no control over pressure or density. To obtain analytical solutions, it's necessary to describe a metric function that relates pressure and density. This article presents a reasonable approach to calculating the metric potential based on Buchdahl ~\cite{Buchdahl:1959zz}. Here we use the metric potential in the following form
\begin{eqnarray}\label{eq3.1}
\sigma=\ln(\zeta(1+x))-\ln(\zeta+x) 
\end{eqnarray}
and
\begin{eqnarray}\label{eq3.2}
\upsilon=2\ln\eta
\end{eqnarray}
where $x =\xi r^{2}$ and  $\eta$ is function of r. The value of $\xi$ is positive and the value $\zeta$ is less than zero. It is not novel to use the Buchdahl ansatz to calculate the metric potential.
Although simple, the model meets the physical limitations of an actual star and maintains regularity and finite conditions at the centre of the sphere. Vaidya and Tikekar ~\cite{Vaidya:1982zz} proposed a metric function where $\xi = \frac{-\zeta}{R^{2}}$. This allowed for a unique geometric interpretation of the model, breaking away from the constraints of 3-space geometry.
Nevertheless, ~\cite{Lake:2002bq} has required that the metric coefficients $ e^{\sigma}$ and $e^{\upsilon}$ be regular and increasing with the increase in r to have a physically valid model. It ought to be positive, devoid of singularity in the centre, with $\upsilon^{\prime}(0)=0$ and $\sigma^{\prime}(0)=0$.\\With the help of Eq.~(\ref{eq2.30}), Eq.~(\ref{eq3.1}) and Eq.~(\ref{eq3.2}), we find the expression for anisotropy factor $\Delta$ given below:   
\begin{equation}\label{eq3.3}
\Delta=\frac{\beta_{1}}{8\pi}\Bigg[\frac{\zeta +x}{\zeta(1+x)}\Bigg[\frac{\eta^{\prime\prime}}{\eta}-\sqrt{\frac{\xi}{x}}\frac{\eta^{\prime}}{\eta}+\frac{\sqrt{\xi x}(\zeta-1)}{(\zeta+x)(1+x)}(\sqrt{\xi x}-\frac{\eta^{\prime}}{\eta})\Bigg]\Bigg].
\end{equation}
Now by using transformation $Y_{1}=\sqrt{\frac{\zeta+x}{\zeta-1}}$ and $\eta=(1-Y_{1}^{2})^{1/4}X$ in Eq.~(\ref{eq3.3}), We get a simpler form of  second order differential equation that is
\begin{equation}\label{eq3.4}
    X^{\prime \prime}+\frac{1}{Y_{1}^{2}-1}\Bigg[1-\zeta +\frac{8\pi \Delta\zeta(1+x)^{2}}{\xi x}+\frac{3Y_{1}^{2}+2}{4(1-Y_{1}^{2})}\Bigg]X=0.
\end{equation}
To obtain the solution of  Eq.~(\ref{eq3.4}), We assume the anisotropy factor $\Delta$ which is described as
\begin{equation}\label{eq3.5} 
\Delta =\frac{\xi x}{8 \pi \zeta(1+x)^{2}}\beta_{1}\Bigg[\frac{5}{4}\frac{1}{(1-Y_{1}^{2})}-\frac{2 \alpha(1-Y_{1}^{2})}{Y_{1}^{2}(\alpha+\beta Y_{1})}+\zeta-\frac{7}{4}\Bigg],
\end{equation}
 where $\alpha$ and $\beta$ are non zero parameters. Putting this value of $\Delta$ in Eq.~(\ref{eq3.4}) we get,
\begin{equation}\label{eq3.6}
    X^{\prime \prime}-\frac{2\alpha}{Y_{1}^{2}(\alpha+\beta Y_{1})}X=0.
\end{equation}
Solving this we get,
\begin{eqnarray}\label{eq3.7}
X= A_{1}\frac{\alpha+\beta Y_{1}}{Y_{1}}\Bigg[\frac{\alpha}{\beta^{3}}f(Y_{1})+\frac{B_{1}}{A_{1}}\Bigg].
\end{eqnarray}
By using this obtained value of X we get $\eta$ as,
\begin{eqnarray}\label{eq3.8}
\eta= A_{1}(1-Y_{1}^{2})^{1/4}\frac{\alpha+\beta Y_{1}}{Y_{1}}\Bigg[\frac{\alpha}{\beta^{3}}f(Y_{1})+\frac{B_{1}}{A_{1}}\Bigg],
\end{eqnarray}
where $f(Y_{1})=\frac{\sec^{2}v-\cos^{2}v}{2}+\log\cos^{2}v$ with $v=\tan^{-1}\sqrt{\frac{\beta Y_{1}}{\alpha}} $, $A_{1}$ and $B_{1}$ are arbitrary  constants and $\zeta<0$.
using Eq.~(\ref{eq3.1}), Eq.~(\ref{eq3.2}) and Eq.~(\ref{eq3.8}) in  Eqs.~(\ref{eq2.27}--\ref{eq2.29}) we get the expression for energy density, radial pressure, and tangential pressure given below:

\begin{eqnarray}\label{eq3.9}
&& \hspace{-0.5cm}\rho=\frac{\beta_{1} \xi (\zeta-1) (3+x)}{8 \pi \zeta(1+ x)^{2}}-\frac{\beta_{2}}{16 \pi},\\
&& \hspace{-0.6cm}
p_{r}=\frac{-2\xi \beta_{1}+\beta_{2}(1+x)+2(1+x)\xi \beta_{1}\psi_{1}}{8\pi 2(1+x)\psi_{2}\psi_{3}}+\frac{2\xi^{2}r^{2}\beta_{1}\psi_{4}}{8\pi\zeta(1+x)^{2}\psi_{5}\psi_{6}}\label{eq3.10},\\
&& \hspace{-0.6cm} p_{t} = \frac{2 B_{1} \beta^{3} (\zeta-1)\psi_{7} + A_{1} \beta \zeta^{2}\beta_{2}\psi_{8} + 2A_{1}(\zeta-1)\alpha \psi_{9}\psi_{10}}{8\pi\psi_{11}\psi_{12}}\label{eq3.11}.
\end{eqnarray}\\
The value of $\psi_{1}$,$\psi_{2}$,$\psi_{3}$,$\psi_{4}$,$\psi_{5}$,$\psi_{6}$,$\psi_{7}$,$\psi_{8}$,$\psi_{9}$,$\psi_{10}$,$\psi_{11}$,$\psi_{12}$ are seen in appendix.

\begin{table*}[!htp]
\centering
\begin{tabular}{cccccccccc}
      \hline
    $\beta_{1}$ & $\beta_{2}$ & $\beta$ & $\alpha$ & $\xi $ & $\zeta$&$\frac{M}{R}$&Compact star&$A_{1}$&$B_{1}$\\
    \hline
    0.8 & 0.00001&0.93& 0.2& 0.002955930 &-0.62&0.1741164& 4U1820-30&2.08102&2.54774 \\
    \hline
    0.9 & 0.00001& 0.93 &0.2&0.002955930&-0.62&0.1741230& 4U1820-30&2.07918&2.54787 \\
    \hline
    1.0 &0  &0.93& 0.2 &0.002955930&-0.62&0.1741747& 4U1820-30&2.06445&2.54893\\
    \hline 
    1.1 &0.00001 & 0.93&0.2 & 0.002955930&-0.62&0.1741318& 4U1820-30&2.07650&2.54807 \\
    \hline
    1.2 &0.00001&0.93&0.2&0.002955930&-0.62&0.1741362& 4U1820-30&2.07550&2.54814 \\ 
    \hline
    \end{tabular}
    \caption{The numerical value of parameters $\alpha$, $\beta$, $\xi$, $\zeta$, $A_{1}$ and $B_{1}$ for different values of $\beta_{1}$ and $\beta_{2}$}
    \label{Table1}
\end{table*}

\section{Necessory and physical existence requirements in $F(\mathcal{Q})$ gravity}\label{sec4}
The following are the necessary and physical existence requirements for the $F(\mathcal{Q})$ gravity model, a well-behaved model must satisfy these requirements:
\begin{itemize}
    \item Singularities must not exist in the solution, meaning that the metric functions, density and pressures must all be non-negative for 0 $\leq$r$\leq R$.
\item The density and pressures ought to be positive, monotonically decreasing toward the surface, with the centre having the largest value.
\item At the border, radial pressure should disappear.
\item The ratio of pressures to the density of the star i:e $\frac{p}{\rho}$ must be positive and have a value greater than zero and less than 1.\\
\begin{eqnarray}\label{eq4.1}
 0<\frac{p}{\rho}<1.   
\end{eqnarray}
\item At the centre of the star, the causality requirements must be satisfied.\\i:e
$0<\Bigg(\frac{dp}{d\rho}\Bigg)_{r=0}\leq1$.\\
\item The value of $\frac{dp}{d\rho} $ must be positive and decrease monotonically as the value of r increases.
\begin{equation}\label{eq4.2}
\Bigg(\frac{d}{dr}\Bigg(\frac{dp}{d\rho}\Bigg)_{r=0}<0.
\end{equation}
\item As the value of r increases the solution value of $\frac{p}{\rho}$ are monotonically decreasing.\\
\begin{equation}\label{eq4.3}
\Bigg(\frac{d}{dr}\Bigg(\frac{p}{\rho}\Bigg)_{r=0}<0.
\end{equation}
  \end{itemize}

\section{Boundary requirements}\label{sec5}
The boundary requirement for solving the Einstein-field equations in $F(\mathcal{Q})$ gravity is presented in this section. In order to accomplish this, we compare the solution on the exterior and interior spacetimes. There are numerous indications in more recent works, such as~\cite{Wang:2021zaz}, that the feasible solution in $F(\mathcal{Q})$ gravity is solvable using Schwarzchild (Anti-) di sitter solution. The solution is provided as follows:

\begin{eqnarray}\label{eq5.1}
ds^{2}=-\Bigg(1-\frac{2M}{r}-\frac{\Lambda}{3}r^{2} \Bigg)dt^{2} + \frac{dr^{2}} {\Bigg(1-\frac{2M}{r}-\frac{\Lambda}{3}r^{2}\Bigg)} \nonumber\\&&\hspace{-7.3cm}+\hspace{0.1cm} r^{2}(d\theta^{2}+\sin^{2}\theta d\phi^{2}).
\end{eqnarray}
Taking $r = R$, the Schwarzchild (Anti-) di sitter metric has been solved at the surface using the requirement that Darmois-Israel derived in their research ~\cite{Israel:1966rt}

\begin{eqnarray}\label{eq5.2}
&& \hspace{-0.5cm}\Bigg(1-\frac{2M}{r}-\frac{\Lambda}{3}r^{2}\Bigg)=e^{\upsilon(R)},\\
&& \hspace{-0.5cm}
\Bigg(1-\frac{2M}{r}-\frac{\Lambda}{3}r^{2}\Bigg)=e^{-\sigma(R)}\label{eq5.3},\\
&& \hspace{1.7cm}p_{r}(R)=0.\label{eq5.4}
\end{eqnarray}

Where the notations $M$ and $\Lambda$ denote the mass and cosmological constant. It is noteworthy that the cosmological constant $\Lambda$ is contingent upon the constants $\beta_{1}$ and $\beta_{2}$, which are expressed as follows $\Lambda =\frac{\beta_{2}}{\beta_{1}}$. In this scenario, the evidence for the cosmological constant is zero, hence its value does not affect the current stellar model. Normally, the cosmological constant would have a significant value to influence the mathematical formulation of the problem. The cosmological constant, which is zero in the current study, is estimated to be $10^{-46}km^{-2}$, based on numerical evidence. For this project, $\beta_{2}$ has been set to 0.00001 for the entire problem.
  Eqs.~(\ref{eq5.2}--\ref{eq5.4}) are boundary conditions with the help of these boundary conditions we find the values of the constants $ A_{1}$ and $B_{1}$  that is

\begin{eqnarray}\label{eq5.5}
&& \hspace{0.9cm} A_{1} = - \frac{2B_{1}(-1+\zeta)\beta^{3}\psi_{13}}{(\zeta+\xi R^{2})\beta \psi_{14} +2(-1+\zeta) \alpha Z},\\
&& \hspace{0.9cm}
B_{1}=\frac{\psi_{15}}{(1+\xi R^{2})(\alpha+Y_{2}\beta)\psi_{16}}\label{eq5.5}. 
\end{eqnarray}
The value of $\psi_{13}$,$\psi_{14}$,$\psi_{15}$, and $\psi_{16}$ are seen in appendix

\section{Physical analysis of the model}\label{sec6}
We conducted several physical experiments using various parameter values to show the physical acceptability of our proposed model. We have included the graphical representations using the parameter values from Table~\ref{Table1} for clarity's sake.

\subsection{Energy density, pressures and anisotropy}\label{sec6.1}
\begin{figure*}[!htp]
    \centering
    \includegraphics[height=7cm,width=7.5cm]{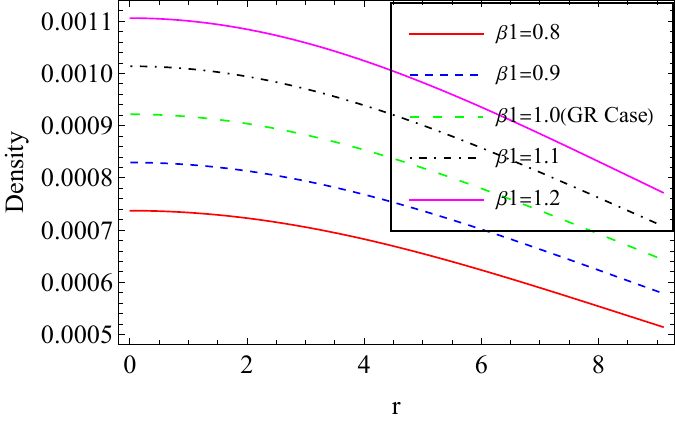}~~~~~~~~~\includegraphics[height=7cm,width=7.5cm]{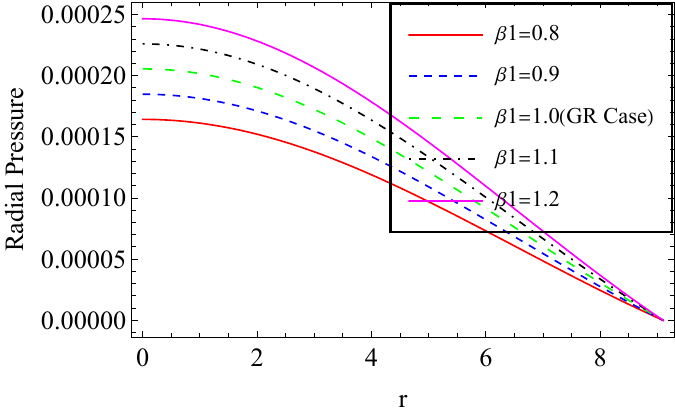}\\~~~~~\includegraphics[height=7cm,width=7.5cm]{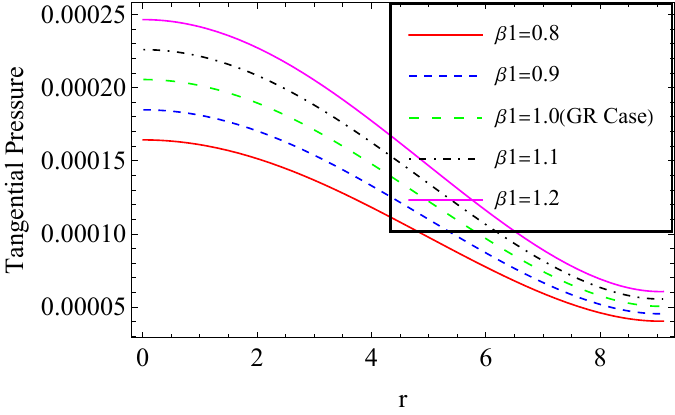}~~~~~\includegraphics[height=7cm,width=7.5cm]{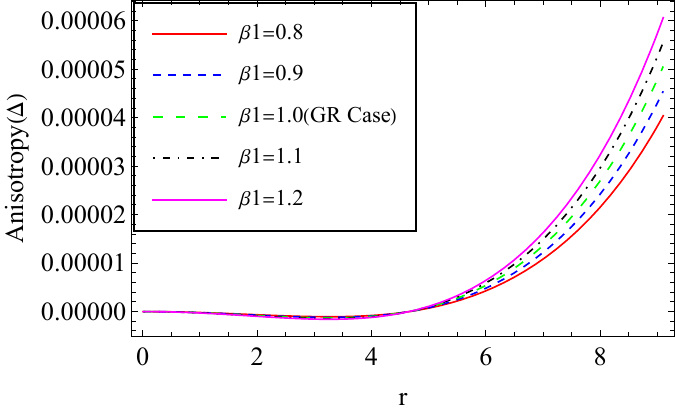}
    \caption{The behaviour of energy density($\rho$), radial pressure($p_{r}$), tangential pressure($p_{t}$) and anisotropy($\Delta$) versus the radius r of the proposed compact star using constant value C = 0.002955930.}
    \label{figure 1}
\end{figure*}

In $F(\mathcal{Q})$ gravity, the pressure and density of a compact star decrease towards the surface, the radial pressure goes to zero on the boundary ($r = R$), and the tangential pressure remains positive throughout the star. It is also noteworthy that the radial and tangential pressures as well as the energy density, are significantly influenced by the parameter $\beta_{1}$.
As the value of $\beta_{1}$ increases, the value of radial and tangential pressures as well as energy density also increases. The Figure~\ref{figure 1} makes this completely evident. The star's interior is more compact than its outer layer, as seen by the behaviour where the density is highest in the centre and decreases towards the surface. Furthermore, we observe that the curves of $p_{r}$ and $p_{t}$ for various $\beta_{1}$ values are monotonically decreasing towards the surface. It follows that the pressure-induced force is positive and exerts an outward force. An essential prerequisite for a stable, realistic stellar object is the presence of a force induced by pressure characteristic that counteracts the gravitational pull to avoid gravitational collapse.
In addition, we have also analysed the celestial body's anisotropy in this section. The two opposing forces affect a celestial body's anisotropy: an attracting force is induced if anisotropy i:e $p_{t}-p_{r}$ is negative and a repulsive force is induced if anisotropy i:e $p_{t}-p_{r}$ is positive. While the matter is drawn towards the body by the attraction force, it is pushed away from it by the repulsive force. At the centre of the celestial body's anisotropy:e $p_{t}-p_{r}$= 0 is zero. There are isotropic conditions, the two forces balance and there is no net force. As one moves out from the compact star's centre, the anisotropy develops monotonically with a positive slope. This rise results from the repulsive force becoming more powerful the further one goes from the compact star. The graph in Figure~\ref{figure 1} illustrates the same. Throughout the star proximity, the anisotropy graph must have a rising behavior and be positive. The graph that is being shown and Table~\ref{Tab2} both show the same behaviour. The mass-radius ratio($\frac{M}{R}$) is shown in Table~\ref{Tab2}, which also provides the mathematical value of density at the centre and surface and the value of pressure at the centre. The compact star exhibits good behaviour, as evidenced by the increasing values of $\beta_{1}$ and the corresponding increases in the mathematical value of density at the centre and surface, pressure at the centre, and mass-radius ratio.

\begin{table*}[!htp]
\centering
\begin{tabular}{ccccc}
      \hline
    $\beta_{1}$ & value of density at centre & value of density at surface & value of pressure at centre &$ \frac{M}{R}$\\
    \hline
    0.8 & 3.95769 $\times10^{13}$&2.76555$\times10^{13}$& 0.792776$\times10^{34}$& 0.1741164 \\
    \hline
    0.9 & 4.45710 $\times10^{13}$& 3.10923$\times10^{13}$ &0.894290$\times10^{34}$&0.1741230\\
    \hline
    1.0 & 4.95114 $\times10^{13}$ &3.45828 $\times10^{13}$& 0.995804 $\times10^{34}$ &0.1741747\\
    \hline 
    1.1 &5.445180$\times10^{13}$ & 3.80196$\times10^{13}$ &1.092484$\times10^{34}$ &0.1741318 \\
    \hline
    1.2 &5.939220$\times10^{13}$&4.14564$\times10^{13}$&1.193998$\times10^{34}$ &0.1741362 \\
    \hline
    \end{tabular}
    \caption{The value of density at centre and surface, and the value of  pressure at the centre for different values of $\beta_{1}$}
    \label{Tab2}
\end{table*}
\subsection{Metric potentials}\label{sec6.2}
\begin{figure*}[!htp]
    \centering
    \includegraphics[height=7cm,width=7.5cm]{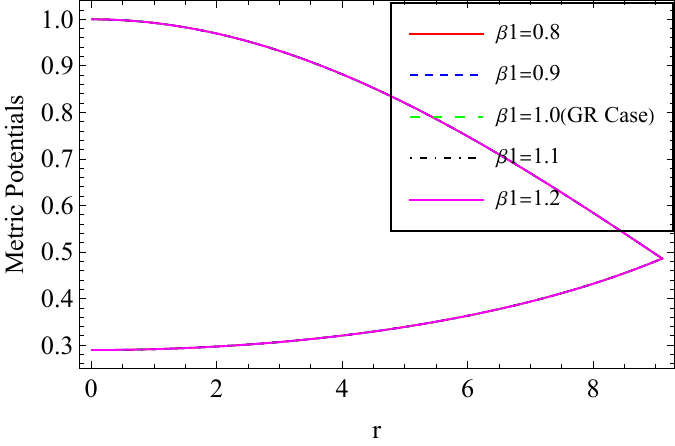}
    \caption{The behaviour of metric potentials $e^{-\sigma} $ and $ e^{\upsilon}$ versus the radius r of the proposed compact star using constant value C = 0.002955930.}
    \label{figure 2}
\end{figure*}
Anisotropic compact stars require finite and nonsingular metric potential functions to produce a physically plausible model. In our model, the value of $e^{\sigma}$ at $r=0$ equals 1, and the value of $e^{\upsilon}$ at r=0 equals 0.04457. we have a positive value of $e^{\sigma}$ and  $e^{\upsilon}$ at r=0 for non-zero $\alpha$ and  $\beta$, indicates that the metric potential is stable at the centre of the compact star. The accurate behaviour of the metric potential potentials has been shown in Figure~\ref{figure 2}. The graph of metric potentials $e^{-\sigma}$ and  $e^{\upsilon}$ are finite and singular throughout the radius of the star and the graph of  $e^{-\sigma}$ is monotonically decreasing towards the surface and the graph of  $e^{\upsilon}$ is monotonically increasing towards the surface from these characteristics of metric potentials we can conclude that metric potentials are competent to produce the model for anisotropic compact star. 

\subsection{Energy state parameter}\label{sec6.3}
\begin{figure*}[!htp]
    \centering
    \includegraphics[height=7cm,width=7.5cm]{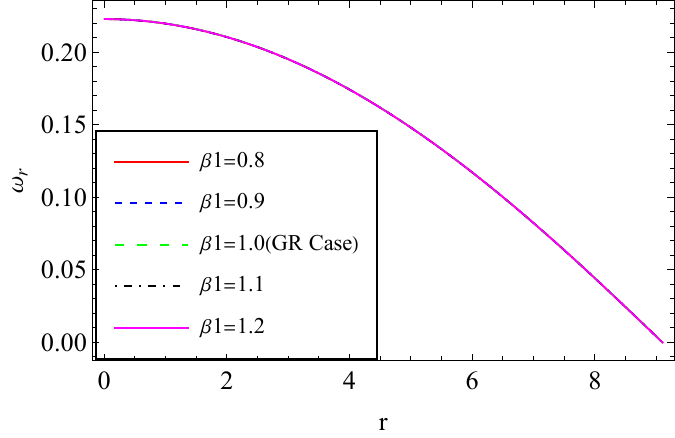}~~~~~\includegraphics[height=7cm,width=7.5cm]{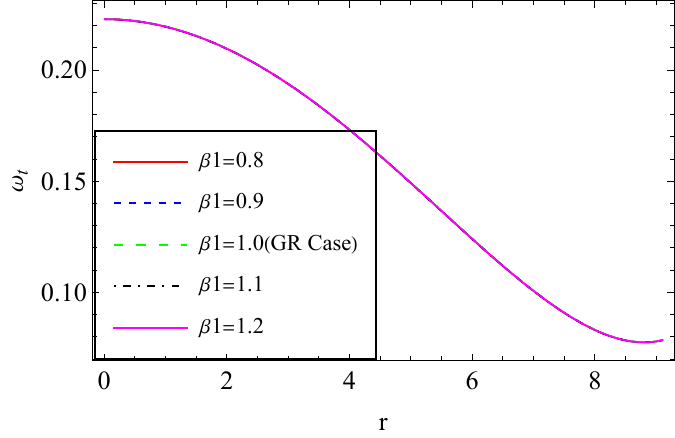} 
    \caption{The behaviour of energy state parameters versus the radius r of the proposed compact star using constant value C = 0.002955930.}
    \label{figure 3}
\end{figure*}

The energy state parameters in radial and transversal form can be shown as: 
\begin{eqnarray}\label{eq6.1}
     \omega_{r}= \frac{p_{r}}{\rho}, 
      \quad \omega_{t}=\frac{p_{t}}{\rho}.
 \end{eqnarray}
  We know that $\omega_{r}$ and $\omega_{t}$ must satisfy the inequality $0<\omega_{r}$, $\omega_{t}<1$ for the model to be viable. Figure~\ref{figure 3} shows a graphical representation of energy state parameters. The energy state parameters have been created and are completely fulfilled. Thus, our model is physically possible. 

\subsection{Stability analysis}\label{sec6.4}
\begin{figure*}[!htp]
    \centering
    \includegraphics[height=7cm,width=7.5cm]{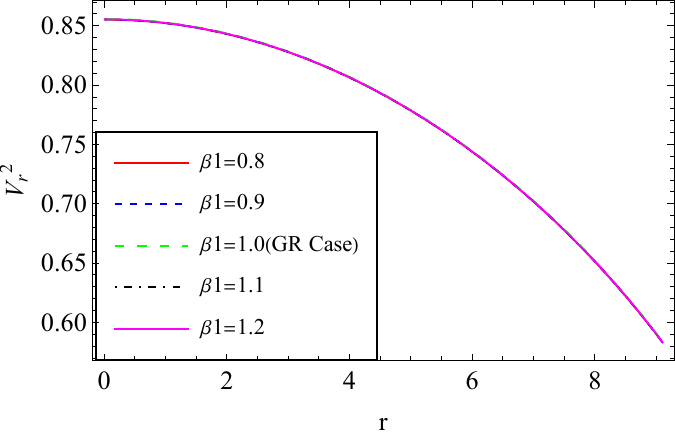}~~~~~\includegraphics[height=7cm,width=7.5cm]{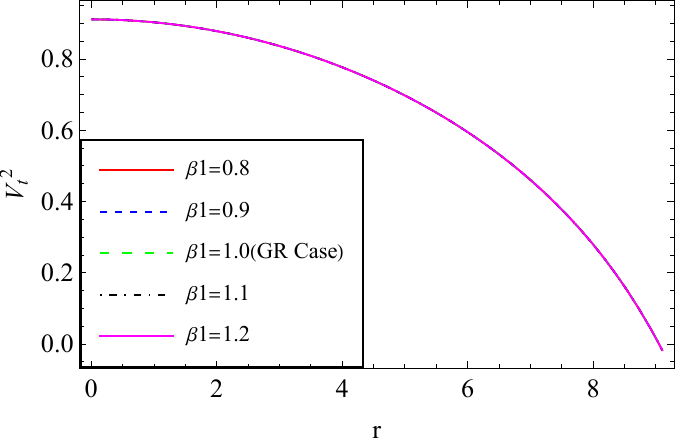} 
    \caption{The behaviour of radial velocity ($V^{2}_{r}$) and Tangential velocity ($ V^{2}_{t} $ ) versus the radius r of the proposed compact star using constant value C = 0.002955930.}
    \label{figure 4}
\end{figure*}
By examining the model for any possible cracks or overturns, we can determine if the star is possibly stable or unstable. At both extremities of the star, the fluid components accelerate to one another.
The stability condition of the compact star explained by Herrera is the possibility of the star splitting at some points within the star, provided that the star does not compress or expand. According to this requirement, the model's radial velocity($V^{2}_{t}$) and tangential velocity($V^{2}_{t}$) must both be smaller than 1. When we look at  Figure~\ref{figure 4}, we can see that the values of $V^{2}_{r}$ and $V^{2}_{t}$ fall between 0 and 1. As a result, the star is steady and Herrera's stability requirement is satisfied.
\begin{eqnarray}\label{eq6.2}
    V^{2}_{r}=\frac{dp_{r}}{d\rho}<1   ,V^{2}_{t}=\frac{dp_{t}}{d\rho}<1,
\end{eqnarray}
Figure~\ref{figure 4} shows a graphical depiction of $V^{2}_{r}$ and $V^{2}_{t}$ and complies with the stability requirements stated by Abreu et al~\cite{Abreu:2007ew}. A stability region was identified by this investigation, when $V^{2}_{r} >V^{2}_{t}$ and the nature of the difference between these velocities must be positive. This shows that the compact star is stable. A generalisation of this claim was proposed by Andreasson~\cite{Andreasson:2007ck, Herrera:1992lwz} who also introduced the idea of no cracking, which indicates the stability of the compact star. The zone that is free of cracks is denoted by $0<|V^{2}_{t}-V^{2}_{r}|<1 $. The illustrations demonstrate the compact star's good mathematical and physical behaviour. We may also examine how $\beta_{1}$ affects on $V^{2}_{t}$ and $V^{2}_{r}$. Figure~\ref{figure 4} shows that the value of $V^{2}_{t}$ and $V^{2}_{r}$ decrease monotonically as the value of $\beta_{1}$ increases.
\subsection{Mass function and the compactness parameter}\label{sec6.5}
\begin{figure*}[!htp]
    \centering
    \includegraphics[height=7cm,width=7.5cm]{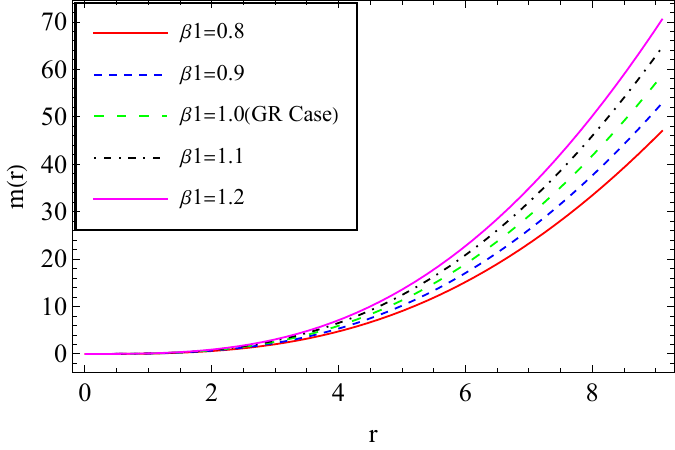}~~~~~\includegraphics[height=7cm,width=7.5cm]{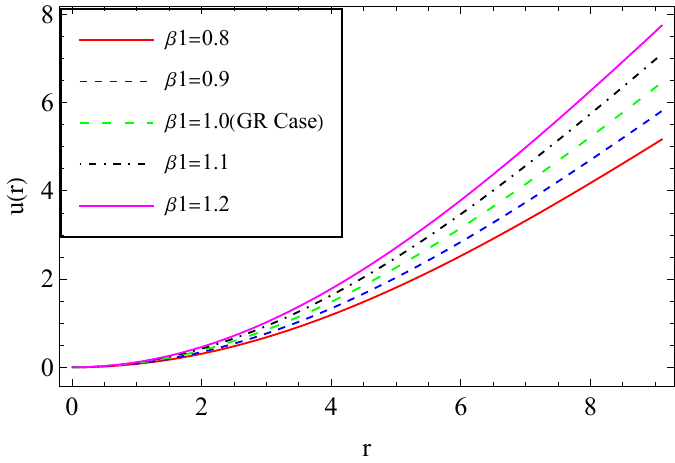}
    \caption{The behaviour of mass function $m(r)$ and compactness parameter $u(r)$ versus the radius r of the proposed compact star using constant value C = 0.002955930.}
    \label{figure 5}
\end{figure*}
The compact star's compactness and mass are shown in Figure~\ref{figure 5}. The impact of parameter $\beta_{1}$ on the compactness and mass of the compact star is visible. When the value of $\beta_{1}$ increases, the value of mass and compactness also increases. The values of mass and compactness are zero at the centre of the compact star, and they monotonically increase towards the surface. It is significant to note that there is a mass-radius ratio bound for any stellar structure, which is provided by Buchdahl and is $\frac{2M}{R} \leqslant \frac{8}{9}$~\cite{Buchdahl:1959zz} This bound has been computed for the compact star in our model and is displayed in Table~\ref{Tab2}. This bound is less than $\frac{4}{9}$, demonstrating that our model satisfies Buchdahl's requirements. The functions $m(r)$, $u(r)$ of radial coordinate r are used to represent the compact star mass and compactness are given below:

\begin{eqnarray}\label{eq6.3}
&& \hspace{-0.5cm}m(r)=\frac{\kappa}{2}\int^{r}_{0}\rho r^{2}dr
,\\ 
&& \hspace{-0.5cm}
u(r)=\frac{\kappa}{2r}\int^{r}_{0}\rho  r^{2}dr.\label{eq6.4}
\end{eqnarray}
The following is the expression for the mass and compactness of the compact star:
\begin{eqnarray}\label{eq6.5}
    && \hspace{-0.5cm}m(r) =-\frac{44 r^3 \left(-6 \text{$\beta $1} \xi (\zeta-1)+\beta_{2} \xi \zeta r^2+\beta_{2}\zeta \right)}{7 \left(\zeta \left(3 \xi r^2+3\right)\right)},\\
&& \hspace{-0.5cm}
u(r)=-\frac{44 r^2 \left(-6 \beta_{1} \xi (\zeta-1)+\beta_{2} \xi \zeta r^2+\beta_{2} \zeta\right)}{7 \left(\zeta \left(3 \xi r^2+3\right)\right)}\label{eq6.6}.
\end{eqnarray}
The star obtained here is inside the star's limit of being a compact star, with a solar mass of 1.58 for a value of $\beta_{1}$=1. The graph's growing tendency suggests that the star's mass rises gradually as the parameter $\beta_{1}$ rises. In this scenario, the constant $\beta_{2}$ is supposed to be 0.00001, therefore it has minimal impact on the mass and compactness of the compact star.

\subsection{Energy requirements}\label{sec6.6}
When analysing data, all energy constraints must be met for the model to be feasible. Here NEC represents the null energy condition, WEC represents the weak energy condition, DEC represents the dominant energy condition, SEC represents the strong energy condition, and TEC depicts the trace energy condition.
\begin{figure*}[!htp]
    \centering
    \includegraphics[height=7cm,width=7.5cm]{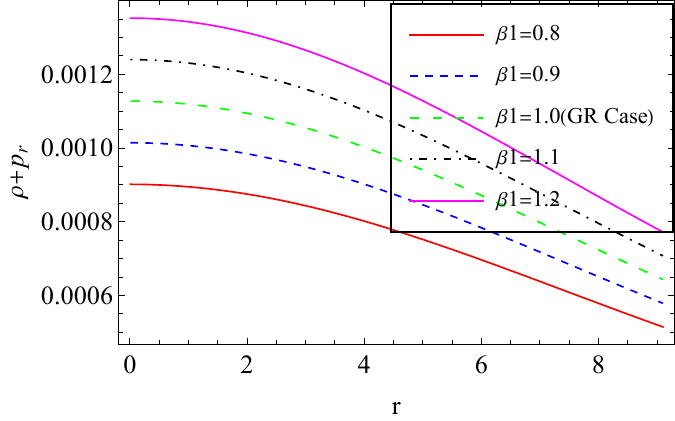}~~~~~\includegraphics[height=7cm,width=7.5cm]{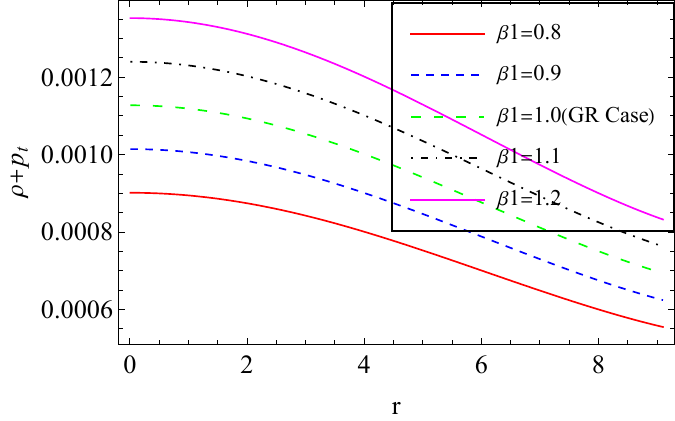}  \\~~~~~\includegraphics[height=7cm,width=7.5cm]{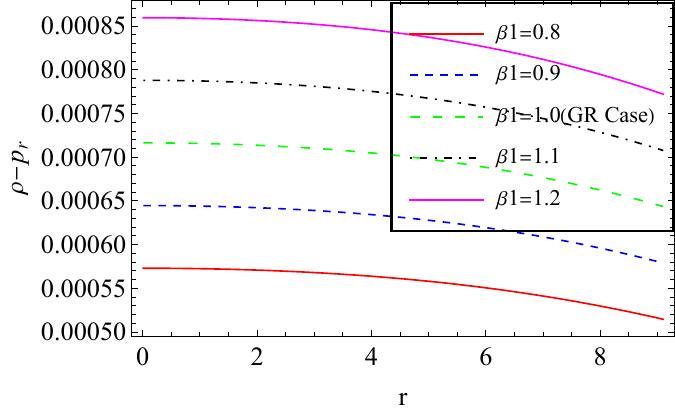}~~~~~\includegraphics[height=7cm,width=7.5cm]{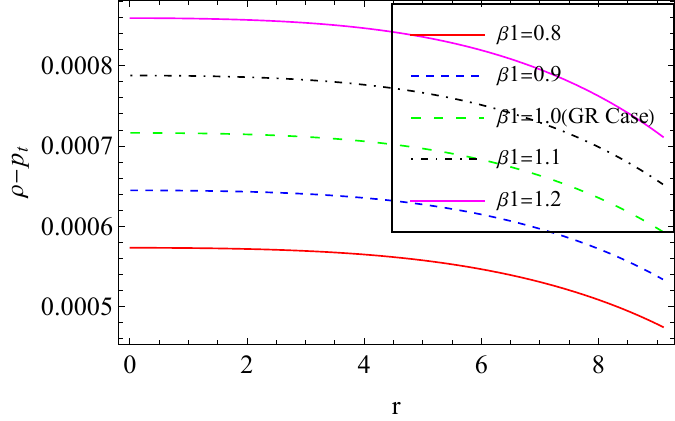}\\~~~~~\includegraphics[height=7cm,width=7.5cm]{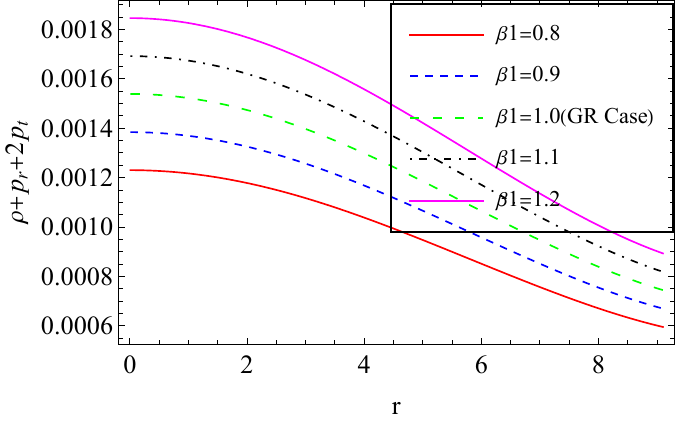}~~~~~\includegraphics[height=7cm,width=7.5cm]{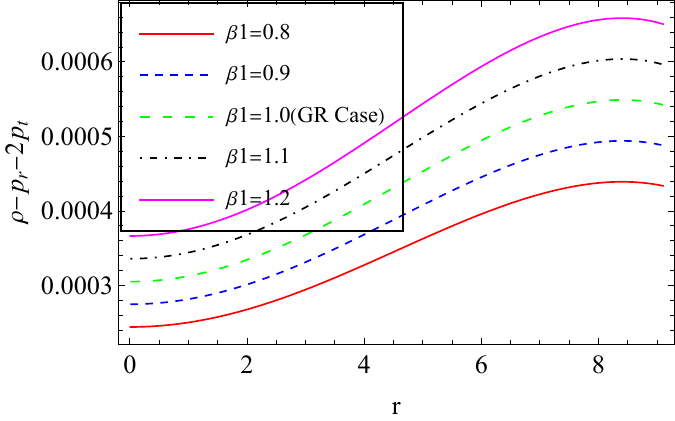} 
    \caption{The behaviour of WEC, DEC, SEC, and TEC  versus the radius r of the proposed compact star using constant value C = 0.002955930.}
    \label{figure 6}
\end{figure*}

\begin{eqnarray}\label{eq6.7}
NEC:\rho \geq 0,\\
WEC:\rho +p_{r}\geq 0,\rho +p_{t}\geq 0,\label{6.8}\\
DEC:\rho -p_{r}\geq 0,\rho -p_{t}\geq 0,\label{6.9}\\
SEC:\rho+p_{r}+2p_{t} \geq 0,\label{6.10}\\
TEC:\rho-p_{r}-2p_{t}\geq0.\label{6.11}
\end{eqnarray}
Figure~\ref{figure 6} represents the energy requirements graphically; all energy requirements needs have been identified and are fully met. Therefore, our solutions are physically possible.

\subsection{Adiabatic index}\label{sec6.7}
\begin{figure}[!htp]
    \centering
    \includegraphics[height=7cm,width=7.5cm]{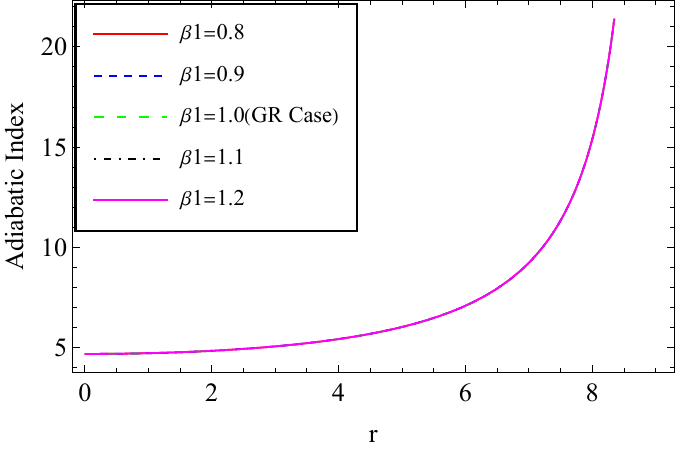}
    \caption{The behaviour of Adiabatic index $\Gamma$  versus the radius r  of the proposed compact star using constant value C = 0.002955930.}
    \label{figure 7}
\end{figure}
One of the most significant stability requirements for each stable structure is the adiabatic index, which we investigate. The adiabatic index can be used to characterize the energy state at a certain energy density. As a result, the adiabatic index influences the stability of compact stars in both relativistic and nonrelativistic systems. Particularly in an anisotropic star, the causality requirement is crucial for ensuring that the model is stable. The adiabatic index needs to be less than $\frac{4}{3}$ for the model to be stable ~\cite{bondi1992anisotropic,heintzmann1975neutron}.The mathematical expression to compute $\Gamma$ is as follows:
\begin{eqnarray}\label{6.12}
\Gamma=\frac{p_{r}+\rho}{p_{r}}.\frac{dp_{r}}{d\rho}
\end{eqnarray}
Bondi's criterion identifies anisotropic gravitational collapse under minor radial perturbations using a compact star's adiabatic index~\cite{bondi1992anisotropic}. A stable Newtonian star requires an average adiabatic index greater than $\frac {4}{3}$. $\Gamma=\frac{4}{3}$ represents a neutral equilibrium for a star's stability. According to Knutsen's~\cite{Pattersons:2021lci} proposal, an isotropic model has an adiabatic index ($\Gamma$) greater than unity if the $\frac{p}{\rho}$ decreases towards the surface, indicating a clear downward trend in temperature. The compact star's adiabatic index  Figure~\ref{figure 5} shows that its value of $\Gamma$ exceeds 4/3. Furthermore, it is evident that altering the value of parameter $\beta_{1}$ does not affect the adiabatic index's trajectory. Thus, $\beta_{1}$ does not influence the adiabatic index. The adiabatic index appears to have a low value at the centre and to be increasing with a minimum value of 4.69, according to the graph in  Figure~\ref{figure 5}. As a result, we can declare that the causality requirement is satisfied.

\subsection{Gravitational and surface  redshift}\label{sec6.8}
\begin{figure*}[!htp]
    \centering
    \includegraphics[height=7cm,width=7.5cm]{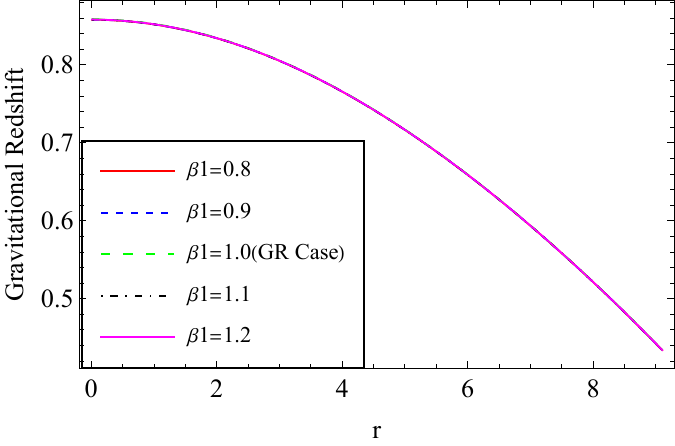}~~~~~\includegraphics[height=7cm,width=7.5cm]{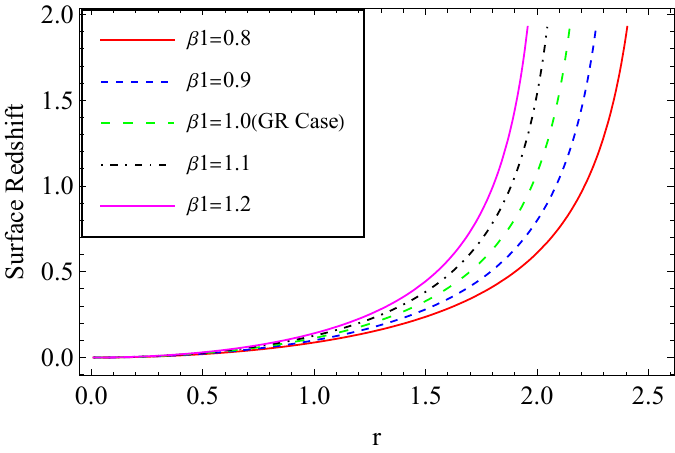}
    \caption{The behaviour of Gravitational Redshift $Z_{G}$ and Surface Redshift $Z_{S}$ versus the radius r of the proposed compact star using constant value C = 0.002955930.}
    \label{figure 8}
\end{figure*}
A phenomenon known as gravitational redshift occurs when electromagnetic waves or photons, lose energy when they emerge from a gravitational well. Electromagnetic radiation expands its wavelength and photon energy to escape when it moves out of a gravitational well. It loses energy not through a change in speed but rather through a change in frequency because it must travel simultaneously at the speed of light. When energy is lost, the photon's wave frequency decreases and its wavelength increases, resulting in a shift toward the red end of the electromagnetic spectrum, commonly known as redshift.. The profile of interior density is similarly presented by the behaviour of interior redshift.
A photon must travel a greater distance and through a denser region the compact star's interior if it is to leave the center and reach the surface. This greater dispersion causes a significant energy loss. A photon that rises close to the surface, will travel a shorter distance through a less dense zone, resulting in less dispersion and energy loss. This means that the interior redshift is highest in the center and lowest toward the surface. The surface redshift depends on the surface gravity, which is defined by the stellar object's radius and total mass. Surface gravity increases as a result of a modest increase in radius, similar to the increase in mass. As a result, the surface redshift increases monotonically towards the surface of star. Gravitational and surface redshifts are defined as:
\begin{eqnarray}\label{6.13}
Z_{G}= \frac{1}{\sqrt{|e^{\upsilon}(r)|}}-1,
\end{eqnarray}
 and the value of $e^{\upsilon}(r)$ is seen in appendix

\begin{eqnarray}\label{6.14}
Z_{S}= \frac{1}{\sqrt{1-2u(r)}} -1 =\nonumber\\&&\hspace{-5.0cm}  \frac{1}{\sqrt{1+2\Bigg(\frac{44 r^2 \left(-6 \beta_{1} \xi (\zeta-1)+\beta_{2} \xi \zeta r^2+\beta_{2} \zeta\right)}{7 \left(\zeta \left(3 \xi r^2+3\right)\right)}}\Bigg)} -1.
\end{eqnarray}

The graph of gravitational redshift is monotonically decreasing towards the surface and the graph of surface redshift is monotonically increasing towards the surface as shown in Figure~\ref{figure 8}.

\subsection{ Tolman-Volkoff-Oppenheimer equation for equilibrium compact objects}\label{sec6.9}
\begin{figure*}[!htp]
    \centering
    \includegraphics[height=8cm,width=11cm]{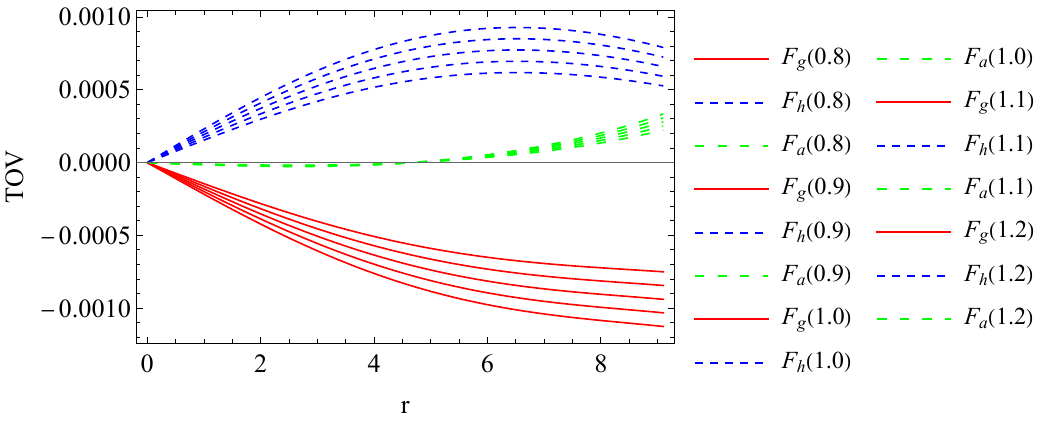}
    \caption{The behaviour of gravitational force ($F_{g}$), hydrostatic force ($ F_{h}$), and anisotropic force ($F_{a}$) versus the radius r of the proposed compact star using constant value C = 0.002955930.}
    \label{figure 9}
\end{figure*}
The equilibrium of an anisotropic stellar object is maintained by the collaborative influence of gravitational force($F_{g}$), hydrostatic force($ F_{h}$) and anisotropic force($F_{a}$). At this stable point, the Tolman-Volkoff-Oppenheimer(TOV) equation~\cite{Tolman:1939jz, Oppenheimer:1938zz} defines the balancing force equation.
\begin{eqnarray}\label{6.15}
F_{g}+F_{h}+F_{a}=0.
\end{eqnarray}
The equation below represents the forces in the TOV equation. 
\begin{eqnarray}\label{6.16}
F_{g}=-\frac{\upsilon^{\prime}}{2}(\rho+p_{r}),F_{h}=-\frac{dp_{r}}{dr},F_{a}=\frac{2}{r}(p_{t}-p_{r}).
\end{eqnarray}
Figure ~\ref{figure 9} illustrates the dominance of ($F_{g}$), which is countered by the collaborative effects of ($ F_{h}$) and ($F_{a}$).This shows that our solutions for anisotropic fluid distributions are in equilibrium. 
\section{Conclusion}\label{sec7}
The study of compact stars has been a fascinating topic of research in astronomy and theoretical physics for the last twenty years. This work introduces the formation of a stable compact star using $F(\mathcal{Q})$ gravity, incorporating a non-perfect fluid distribution of matter. It is crucial to highlight that the formation of a compact star under $F(\mathcal{Q})$ gravity behaves equivalently to that under GR. As a result, any solution derived in GR can also be obtained using $F(\mathcal{Q})$ gravity. To derive stable results, we utilised the reliable Buchdhal metric potential to produce a different solution within the framework of $F(\mathcal{Q})$ gravity theory. The preceding section provides a thorough explanation and investigation of the conclusions of computations conducted within the framework of $F(\mathcal{Q})$ gravity. Various values of  $\beta_{1}$ and $\beta_{2}$  were explored and their impacts on the numerical outcomes of the model were examined as depicted in the graphs presented earlier. The model's thermodynamic quantities have been investigated in detail, and the figures presented the findings of this analysis. Figure~\ref{figure 1} illustrates the variations in thermodynamic quantities of the model corresponding to the constant C = 0.002955930.  Table~\ref{Tab2} provides the numerical values of the thermodynamic quantities derived from these observations. It's important to recognize that the thermodynamic properties vary with the parameter $\beta_{1}$, which was tested at values of 0.8, 0.9, 1.0, 1.1, and 1.2. As $\beta_{1}$ increases, there is a consistent increase in these properties. The figures show smooth behaviour and all necessary requirements are satisfied. In this work, we examined  $V^{2}_{r}$, $V^{2}_{t}$, and $\frac{p}{\rho}$ of the model under consideration. Figure~\ref{figure 3} and Figure.~\ref{figure 4} depict these relationships graphically for the constant C = 0.002955930. Graphs for all these quantities were plotted using varying values of $\beta_{1}$ as previously listed. As $\beta_{1}$ increased, there was a consistent upward trend in the values depicted on the graphs. These quantities vary directly with $\beta_{1}$. The trajectories of $V^{2}_{r}$ and $V^{2}_{t}$ decrease monotonically, ensuring that neither velocity exceeds the speed of light at any point, which fulfils the causality requirement. The pressure density ratio, specifically the ratio of radial pressure to density i:e $\frac{p_{r}}{\rho}$, exhibits a decreasing trend in its graph. For the ratio of tangential pressure to density i:e $\frac{p_{t}}{\rho}$, the graph initially decreases until $r = 8.82$ and then shows a gradual increase. Furthermore, the graphs exhibited smooth behaviour, indicating stability in the model during this specific analysis. Figure~\ref{figure 6} examines the energy requirements of a compact star. The positive trends observed in the thermodynamic quantities depicted in Figure~\ref{figure 1} declare the validity of the SEC, WEC, NEC, and DEC for the compact star. The graph of DEC is plotted in  Figure~\ref{figure 6}, demonstrating the stable characteristics of the compact star. Both DEC and TEC maintain positive values across the region surrounding the compact star. Figure~\ref{figure 5}, Figure~\ref{figure 7} and Figure~\ref{figure 8} illustrate the Mass, Adiabatic index, and redshift of the compact star. As the values of $\beta_{1}$ increase, there is a corresponding increase in the mass of the compact star. The adiabatic index shows a rising trend as compact star mass. If the value of $\beta_{1}$ rises, the gravitational redshift graph consistently decreases throughout the vicinity of the star and the graph of surface redshift is increasing towards the surface. Figure~\ref{figure 9} presents a visual depiction of various forces within the compact star model. The graph confirms the satisfaction of the TOV equation. $F_{g}$, acts inwards, suggesting attractiveness in nature. Conversely, $F_{a}$ and $F_{h}$ exert outward pressure, suggesting they are repulsive in nature. Importantly, the sum of these forces is zero, expressed as  $F_{g}$ + $F_{h}$ + $F_{a}$  = 0. In summary, this work introduces a formation of compact stars within $F(\mathcal{Q})$ gravity, focusing on the study of non-perfect fluid distribution. It is noted that the formation in $F(\mathcal{Q})$ gravity yields identical results to those in GR, implying that solutions obtained in GR are also obtained within the framework of $F(\mathcal{Q})$ gravity. By using the Buchdahl metric, the study produced a new solution within the $F(\mathcal{Q})$ gravity theory, ensuring the attainment of stable solutions. Overall, $F(\mathcal{Q})$ gravity is useful for producing anisotropic solutions in GR and modified gravity theories.
\section*{Acknowledgments}
 Sat Paul acknowledges the Ministry of Tribal Affairs, New Delhi, India, for National Fellowship and Scholarship for Higher Education of ST Students (Award No-202324-NFST-JAM-02473). Jitendra Kumar acknowledges the authority of the Central University of Haryana (CUH). S. K. Maurya thanks the administration of the University of Nizwa in the Sultanate of Oman for their constant support and encouragement. I am very thankful to Sourav Chaudhary a research scholar at CUH for providing study materials for this work. I am highly grateful to Sweeti Kiroriwal from CUH for advising me occasionally regarding my work, which enabled me to write this research paper.
 \section*{Data access statement}
This manuscript includes no accompanying data, therefore the data will not be deposited. The essential calculations and graphic description are already included in the paper.

 \section*{Conflict of interest}
In this research paper, The authors have no conflicts of interest.

\section*{Appendix}
\begin{eqnarray*}
&&\hspace{-0.6cm}Y_{1}= \sqrt{\frac{\zeta+x}{\zeta-1}},\quad Y_{2}= \sqrt{\frac{\zeta+\xi R^{2}}{\zeta-1}},\quad Y_{3}= \sqrt{\frac{\zeta+\xi R^{2}}{\zeta+\xi \zeta R^{2}}}\\\\
&&\hspace{-0.7cm}\psi_{1}=2 A_{1} \beta ^3 Y_{1}+\frac{2 \beta }{(\zeta-1) (x+1) Y_{1} (\alpha +\beta  Y_{1})}-\frac{2}{(x+1) (\zeta+x)}+\frac{1}{\zeta x+\zeta}+\frac{4}{(x+1)^2},\quad \psi_{2} =(1+x)(\alpha+ Y_{1}\beta)\\\\
&& \hspace{-0.7cm}\psi_{3}= \beta \Bigg(2A_{1}(-1+\zeta)  Y_{1} \alpha+A_{1}(\zeta+x)\beta+ 2 B_{1}(-1+\zeta)\beta^{2}(\alpha+ Y_{1}\beta)\Bigg)+ 2A_{1}(-1+\zeta)\alpha(\alpha+ Y_{1}\beta)\log[\frac{\alpha}{\alpha+ Y_{1}\beta}]\\\\
&& \hspace{-0.7cm} \psi_{4}=  A_{1} \beta ^3 (x+1)  Y_{1}+\frac{\beta  (x+1)}{(\zeta-1)  Y_{1} (\alpha +\beta   Y_{1})}-\frac{x+1}{\zeta+x}+2,\quad \psi_{5}=\alpha +\beta Y_{1}\\\\
&& \hspace{-0.7cm}\psi_{6}=\beta  \left( A_{1} \beta  (\zeta+x)+2 \alpha  A_{1} (\zeta-1)  Y_{1}+2 \beta ^2  B_{1} (\zeta-1) (\alpha +\beta  Y_{1})\right)+2 \alpha   A_{1} (\zeta-1) (\alpha +\beta  \text{Y1}) \log \left(\frac{\alpha }{\alpha +\beta  Y_{1}}\right)\\\\ 
  && \hspace{-0.7cm}\psi_{7}=\xi \zeta \psi_{74} +\beta ^2 \xi x^4 \psi_{72}+\xi x^3\psi_{73}+\xi x^2 \psi_{75}+\xi x \psi_{76}+\text{$\beta $2} \zeta^2 \psi_{71}\\\\
  && \hspace{-0.7cm}\psi_{71}=\alpha ^3 (\zeta-1)+3 \alpha  \beta ^2 \zeta+3 \alpha ^2 \beta  (\zeta-1) Y_{1}+\beta ^3 \zeta Y_{1},\quad \psi_{72}=3 \alpha  \left(6 \beta_{1}+\beta_{2} \zeta r^2\right)+\beta  Y_{1} \left(8 \beta_{1}+\beta_{2} \zeta r^2\right)\\\\
 && \hspace{-0.7cm}\psi_{73}=\alpha ^3 (\zeta-1) \left(2 \beta_{1}+\beta_{1} \zeta r^2\right)+\alpha  \beta ^2 \left(\beta_{1} (38 \zeta+36) +3 \beta_{2} \zeta (2 \zeta+3) r^2\right)+3 \alpha ^2 \beta  (\zeta-1) Y_{1} \left(4 \beta_{1}+\beta_{2} \zeta r^2\right)\nonumber\\&&\hspace{-0.1cm} +\beta ^3 Y_{1} \left(2 \beta_{1} (9 \zeta+7)+\beta_{2} \zeta (2 \zeta+3) r^2\right)\\\\
  &&\hspace{-0.7cm} \psi_{74} =\alpha ^3 (\zeta-1) \left(\beta_{1} (6 \zeta-2)+\beta_{1} (3 \zeta+1) r^2\right)+\alpha \beta ^2 \zeta \left(2 \beta_{1}(9 \zeta+1) +3 \beta_{2} (3 \zeta+2) r^2\right)\nonumber\\&&\hspace{-0.3cm}+\alpha ^2 \beta  (\zeta-1)  Y_{1} \left(2 \beta_{1} (9 \zeta-1)+3 \beta_{2} (3 \zeta+1) r^2\right)+\beta ^3 \zeta  Y_{1} \left(\beta_{1} (6 \zeta+2) +\beta_{2} (3 \zeta +2) r^2\right)\\\\
  &&\hspace{-0.7cm}\psi_{75}=\alpha  \beta ^2 \left(\beta_{1} \left(26 \zeta^2+82 \zeta+6\right)+3 \beta_{2} \zeta \left(\zeta^2+6 \zeta+3\right) r^2\right)+\alpha ^2 \beta  \left(\zeta^2+2 \zeta-3\right) Y_{1} \left(10 \beta_{1}+3 \beta_{2} \zeta r^2\right)\nonumber\\&&\hspace{0.1cm}+\beta ^3 Y_{1} \left(2 \beta_{1} \left(6 \zeta^2+17 \zeta+1\right)+\beta_{2} \zeta \left(\zeta^2+6 \zeta+3\right) r^2\right)+\alpha ^3 (\zeta-1) \left(8 \beta_{1}+\beta_{2} \zeta (\zeta+3) r^2\right)\\
  \\&& \hspace{-0.7cm} \psi_{76}= \alpha ^3 (\zeta-1) \left(2\beta_{1} \left(\zeta^2+3 \zeta+1\right)+3 \beta_{1} \zeta (\zeta+1) r^2\right)+\alpha  \beta ^2 \zeta \left(\beta_{1} \left(6 \zeta\zeta^2+64 \zeta+8\right)+3 \beta_{2} \left(3 \zeta^2+6 \zeta+1\right) r^2\right)\nonumber\\&&\hspace{0.1cm}+\alpha ^2 \beta  (\zeta-1) Y_{1} \left(\beta_{1} \left(6 \zeta^2+32 \zeta+6\right)+9\beta_{2} \zeta (\zeta+1) r^2\right)+\beta ^3 \zeta Y_{1}\left(2  \beta_{1} \left(\zeta^2+13 \zeta+2\right)+\beta_{2} \left(3 \zeta^2+6 \zeta+1\right) r^2\right)\\\\
  && \hspace{-0.7cm} \psi_{8}=\psi_{81}+\xi x^{5}\beta^{3}(18\beta_{1}+\zeta r^{2}\beta_{2})+\xi x^{4}\beta\psi_{82}+\xi\zeta\psi_{83}+\xi x^{3}\psi_{84}+\xi x\psi_{85}+\xi x^{2}\psi_{86}\\\\
   && \hspace{-0.7cm} \psi_{81}= \beta ^3 \zeta^2+5 \alpha ^2 \beta  (\zeta-1) \zeta+2 \alpha ^3 (\zeta-1)^2 Y_{1}+4 \alpha  \beta ^2 (\zeta-1) \zeta Y_{1}\\\\
   && \hspace{-0.7cm} \psi_{82}=\alpha ^2 (\zeta-1) \left(22\beta_{1}+5 \beta_{2}\zeta r^2\right)+\beta ^2 \left(28 \beta_{1} (2 \zeta+1)+3 \beta_{1}\zeta (\zeta+1) r^2\right)+4 \alpha  \beta  (\zeta-1) Y_{1} \left(9 \beta_{1}+\beta_{2} \zeta r^2\right)\\\\
    && \hspace{-0.7cm} \psi_{83} = 3 \beta ^3 \zeta^2 (\zeta +1) \left(2 \beta_{1} +\beta_{2} r^2\right)+ \alpha ^2 \beta  (\zeta-1) \zeta \left(\beta_{1} (30 \zeta-2)\nonumber + 5 \beta_{2} (3 \zeta +2) r^2\right)\nonumber\\&&\hspace{0.1cm}+2 \alpha ^3 (\zeta-1)^2 Y_{1} \left(\beta_{1} (6 \zeta-2) + \beta_{2} (3 \zeta+1) r^2\right)+4 \alpha  \beta ^2 (\zeta-1) \zeta Y_{1} \left(\beta_{1} (6 \zeta + 2) + \beta_{2} (3 \zeta + 2) r^2\right)\\\\
     && \hspace{-0.7cm} \psi_{84}=3 \beta ^3 \left(\beta_{1} \left(20 \zeta^2+30 \zeta+2\right)+\beta_{2} \zeta \left(\zeta^2+3 \zeta+1\right)r^2\right)+\alpha ^2 \beta  (\zeta-1) \left(\beta_{1} (42 \zeta+52)+5\beta_{2} \zeta (2 \zeta+3) r^2\right)\nonumber\\&&\hspace{0.1cm}+2 \alpha ^3 (\zeta-1)^2 Y_{1} \left(2 \beta_{1}+\beta_{2} \zeta r^2\right)+4 \alpha  \beta ^2 (\zeta-1)Y_{1}\left(\beta_{1} (19 \zeta+16)+\beta_{2} \zeta (2 \zeta+3) r^2\right)
   \end{eqnarray*}
  \begin{eqnarray*}
   &&\hspace{-0.8cm}\psi_{85} = \alpha^2 \beta  (\zeta-1)\zeta\left(2 \beta_{1} \left(5 \zeta^2+44 \zeta+4\right)+5 \beta_{2} \left(3 \zeta^2+6 \zeta+1\right) r^2\right)\nonumber\\&&\hspace{0.1cm}+\beta ^3 \zeta^2 \left(2 \beta_{1} \left(\zeta^2+23 \zeta+9\right)+3 \beta_{2} \left(\zeta^2+3 \zeta+1\right) r^2\right)\nonumber\\&&\hspace{0.1cm}+2 \alpha ^3 (\zeta-1)^2 Y_{1}\left(2\beta_{1}\left(\zeta^2+3 \zeta+1\right)+3\beta_{2} \zeta (\zeta+1) r^2\right)\nonumber\\&&\hspace{0.1cm}+4 \alpha \beta ^2 (\zeta-1) \zeta Y_{1}\left(\beta_{1}\left (2\zeta^2+26 \zeta+5\right)+\beta_{2} \left(3 \zeta ^2+6 \zeta+1\right)r^2 \right)\\\\
   &&\hspace{-0.8cm}\psi_{86}= 5\alpha ^{2} \beta (\zeta-1) \left(\beta_{1} \left(6 \zeta^{2}+22 \zeta+2\right)+\beta_{2} \zeta \left(\zeta^{2}+6 \zeta+3\right) r^{2}\right)\nonumber\\&&\hspace{0.1cm}+4 \alpha  \beta ^{2} (\zeta-1) Y_{1} \left(3 \beta_{1} \left(4 \zeta^{2}+12 \zeta+1\right) + \beta_{2}\zeta \left(\zeta^{2}+6 \zeta+3\right) r^{2}\right))\nonumber\\&&\hspace{0.1cm}+\beta ^{3} \zeta \left(6 \beta_{1} \left(4 \zeta^{2}+17 \zeta+3\right)+\beta_{2} \left(\zeta^{3}+9 \zeta^{2}+9 \zeta+1\right) r^{2}\right)+2 \alpha ^{3} (\zeta-1)^{2} Y_{1} \left(8  \beta_{1}+\beta_{2} \zeta (\zeta+3) r^{2}\right)\\\\
&&\hspace{-0.8cm}\psi_{9}=\zeta^{2}\beta{2}\psi_{91}+\xi x^{4}\beta^{2}\psi_{92}+\xi x^{3}\psi_{93}+\xi \zeta\psi_{94}+\xi x^{2}\psi_{95}+\xi x\psi_{96}\\\\
&&\hspace{-0.8cm}\psi_{91}= \alpha ^3 (\zeta-1)+3 \alpha\beta ^2 \zeta+3 \alpha ^2 \beta  (\zeta-1) Y_{1}+\beta ^3 \zeta Y_{1},\quad \psi_{92}= 3 \alpha  \left(6 \beta  _{1} +\beta_{2} \zeta r^2\right)+\beta  Y_{1} \left(8 \beta_{1}+\beta_{2} \zeta r^2\right)\\\\
  && \hspace{-0.8cm} \psi_{93}=\alpha ^3 (\zeta-1) \left(2 \beta_{1}+\beta_{2} \zeta r^2\right)+\alpha  \beta ^2 \left(\beta_{1} (38 \zeta+36)+3 \beta_{2} \zeta (2 \zeta+3) r^2\right)+3 \alpha ^2 \beta  (\zeta-1) Y1_{1}\left(4 \beta_{1}+\beta_{1} \zeta r^2\right)\nonumber\\&&\hspace{0.1cm}+\beta ^3 Y_{1} \left(2 \beta_{1} (9 \zeta+7)+\beta_{2} \zeta (2 \zeta+3) r^2\right)\\\\
 && \hspace{-0.8cm} \psi_{94}=\alpha ^3 (\zeta-1) \left(\beta_{1} (6 \zeta-2)+\beta_{2}(3 \zeta+1) r^2\right)+\alpha  \beta ^2 \zeta \left(2 \beta_{1} (9 \zeta+1)+3 \beta_{1} (3 \zeta+2) r^2\right)\nonumber\\&&\hspace{0.1cm}+\alpha ^2 \beta  (\zeta-1) Y_{1} \left(2 \beta_{1} (9 \zeta-1)+3 \beta_{2} (3\zeta+1) r^2\right)+\beta ^3\zeta Y_{1} \left(\beta_{1} (6 \zeta+2)+\beta_{2} (3 \zeta+2) r^2\right)\\\\
 && \hspace{-0.8cm} \psi_{95}=\alpha  \beta ^2 \left(\beta_{1} \left(26 \zeta^2+82 \zeta+6\right)+3 \beta_{2} \zeta \left(\zeta^2+6 \zeta+3\right) r^2\right)+\alpha ^2 \beta  \left(\zeta^2+2 \zeta-3\right) Y_{1} \left(10 \beta_{1}+3 \beta_{2} \zeta r^2\right)\nonumber\\&&\hspace{0.1cm}+\beta ^3 Y_{1} \beta_{1}\left(6 \zeta^2+17 \zeta+1\right)+\beta_{2} \zeta \left(\zeta^2+6 \zeta+3\right) r^2)+\alpha ^3 (\zeta-1) \left(8\beta_{1}+\beta_{2} \zeta (\zeta+3) r^2\right)\\\\
 && \hspace{-0.8cm} \psi_{96}=\alpha ^3 (\zeta-1) \left(2\beta_{1} \left(\zeta^2+3 \zeta+1\right)+3 \beta_{2}\zeta (\zeta+1) r^2\right)+\alpha  \beta ^2 \zeta \left(\beta_{1} \left(6 \zeta^2+64 \zeta+8\right)+3 \beta_{2} \left(3 \zeta^2+6 \zeta+1\right) r^2\right)\nonumber\\&&\hspace{0.1cm}+\alpha ^2 \beta  (\zeta-1)Y_{1}\left(\beta_{1} \left(6 \zeta^2+32 \zeta+6\right)+9 \beta_{2} \zeta (\zeta+1) r^2\right)+\beta ^3 \zeta Y_{1} \left(2 \beta_{1} \left(\zeta^2+13 \zeta+2\right)+\beta_{2} \left(3 \zeta^2+6 \zeta+1\right) r^2\right)\\\\
 && \hspace{-0.8cm} \psi_{10}=\log \left(\frac{\alpha }{\alpha +\beta  Y_{1}}\right),\quad \psi_{11}= 2 (\zeta-1) \zeta (x+1)^3 (\zeta+x) (\alpha +\beta  Y_{1})^2\\\\
  && \hspace{-0.8cm} \psi_{12}=\beta  \left( A_{1} \beta  (\zeta+x)+2 \alpha   A_{1}(\zeta-1)  Y_{1}+2 \beta ^2  B_{1} (\zeta-1) (\alpha +\beta   Y_{1})\right)+2 \alpha   A_{1}(\zeta-1) (\alpha +\beta   Y_{1}) \log \left(\frac{\alpha }{\alpha +\beta   Y_{1}}\right)\\\\
  &&\hspace{-0.8cm}\psi_{13}=\Bigg(\zeta \beta_{2} \psi_{131}+\xi^{3}R^{4}\beta \psi_{132}+\xi^{2}R^{2}\psi_{133}+\xi\psi_{134}\Bigg)\\\\
  && \hspace{-0.8cm} \psi_{131}=2 \alpha  \beta  \zeta+\beta ^2 \zeta Y_{2}+(\zeta-1) Y_{2}\alpha^2,\quad \psi_{132}=\alpha  \left(2 \beta_{2} \zeta R^2-4\beta_{1} (\zeta-4)\right)+\beta Y_{2} \left(\beta_{2} \zeta R^2-2 \beta_{1} (\zeta-5)\right)\\\\
   && \hspace{-0.8cm} \psi_{133}=\beta ^2 Y_{2} \left(\beta_{2} \zeta (\zeta+2) R^2-2 \beta_{1} \left(\zeta^2-8 \zeta-1\right)\right)+2 \alpha  \beta  \zeta \left(\beta_{2} (\zeta+2) R^2-2 \beta_{1} (\zeta-7)\right)\nonumber\\&&\hspace{-0.0cm}+\alpha ^2 (\zeta-1) Y_{2} \left(\beta_{2} \zeta R^2-2 \beta_{1} (\zeta-3)\right)\\\\
   && \hspace{-0.8cm} \psi_{134}=2 \alpha  \beta  \zeta \left(6 \beta_{1} \zeta+2 \beta_{2} \zeta R^2+\beta_{2} R^2\right)+2 \alpha ^2 (\zeta-1) Y_{2} \left(\beta_{1} (3 \zeta-1)+\beta_{2} \zeta R^2\right)\nonumber\\&&\hspace{-0.0cm}+\beta ^2 \zeta Y_{2} \left(\beta_{1} (6 \zeta+2)+\beta_{2} (2 \zeta+1) R^2\right)\\\\
   && \hspace{-0.8cm} \psi_{14}=\Bigg(\zeta\beta{2}\psi_{141}+\xi^{3}R^{4}\beta^{2}(-2(-7+\zeta)\beta{1}+\zeta\beta^{2}\beta{2})+\xi^{2}R^{2}\psi_{142}+\xi\psi_{143}\Bigg)
\end{eqnarray*}
 \begin{eqnarray*}
&& \hspace{-0.8cm}\psi_{141}=2 \alpha  \beta  \zeta+\alpha ^2 (\zeta-1) Y_{2}+\beta ^2 \zeta Y_{2}\\\\  
&& \hspace{-0.8cm}\psi_{142}=\beta ^2 \left(\beta_{1} \left(-2 \zeta^2+20 \zeta+6\right)+\beta_{2} \zeta (\zeta+2) R^2\right)-2 \alpha ^2 (\zeta-1) \left(2 \beta_{1} (\zeta-3)-\beta_{2} \zeta R^2\right)\nonumber\\&&\hspace{-0.0cm}+\alpha  \beta  (\zeta-1) Y_{2} \left(\beta_{1} (26-6 \zeta)+3 \beta_{2} \zeta R^2\right)\\\\
&& \hspace{-0.8cm}\psi_{143}= 4 \alpha ^2 (\zeta-1) \left(\beta_{1} (3 \zeta-1)+\beta_{2} \zeta R^2\right)+\beta ^2 \zeta \left(6 \beta_{1} (\zeta+1)+\beta_{2} (2 \zeta+1) R^2\right)\nonumber\\&&\hspace{-0.0cm}+2 \alpha  \beta  (\zeta-1)Y_{2}\left(\beta_{1}+9 \beta_{1} \zeta+3 \beta_{2} \zeta R^2\right)\\\\
&&\hspace{-0.8cm}Z=\Bigg(\zeta\beta{2}Z_{1}+\xi^{3}R^{4}\beta Z_{2}+\xi^{2}R^{2}Z_{3}+\xi Z_{4}\Bigg)\Bigg(\log\Bigg[\frac{\alpha}{\alpha+Y2}\Bigg]\Bigg),\quad Z_{1}=2 \alpha  \beta  \zeta+\alpha ^2 (\zeta-1) Y_{2}+\beta ^2 \zeta Y_{2}\\\\
 && \hspace{-0.8cm}Z_{2}= \alpha  \left(2 \beta_{2} \zeta R^2-4 \beta_{1} (\zeta-4)\right)+\beta  Y_{2} \left(\beta_{2} \zeta R^2-2 \beta_{1} (\zeta-5)\right)\\\\ 
&& \hspace{-0.8cm}Z_{3}=\beta ^2 Y_{2} \left(\beta_{2} \zeta (\zeta+2) R^2-2 \beta_{1} \left(\zeta^2-8 \zeta-1\right)\right)+2 \alpha  \beta  \zeta \left(\beta_{2} (\zeta+2) R^2-2 \beta_{1} (\zeta-7)\right)\nonumber\\&&\hspace{-0.2cm}+\alpha ^2 (\zeta-1)Y_{2} \left(\beta_{2} \zeta R^2-2 \beta_{1} (\zeta-3)\right)\\\\
&& \hspace{-0.8cm}Z_{4}=2 \alpha  \beta  \zeta \left(6 \beta_{1} \zeta+2 \beta_{2} \zeta R^2+\beta_{2} R^2\right)\nonumber\\&&\hspace{-0.2cm}+2 \alpha ^2 (\zeta-1) Y_{2} \left(\text{$\beta $1} (3 \zeta-1)+\text{$\beta $2} \zeta R^2\right)+\beta ^2 \zeta Y_{2} \left(\beta_{1} (6 \zeta+2)+\beta_{2} (2 \zeta+1) R^2\right)\\\\
&& \hspace{-0.8cm}\psi_{15}=4 (1-\zeta) Y_{2} Y_{3},\quad {\psi_{16}}=\Bigg(1-\frac{\psi_{17}}{\psi_{18}}\Bigg)\\\\
&& \hspace{-0.8cm}{\psi_{17}} = 2(-1+\zeta)\alpha\frac{1}{2}\Bigg(1+\frac{Y_{2}\beta}{2}-\frac{\alpha}{\alpha+Y_{2}\beta}\Bigg)\nonumber\\&&\hspace{-0.2cm}+2(-1+\zeta)\log\Bigg[\frac{\alpha}{\alpha+Y_{2}\beta}\Bigg]\Bigg(\zeta\beta_{2} \psi_{171}+\xi^{3}R^{4}\beta\psi_{172}+\xi^{2}R^{2}\psi_{173}+\xi\psi_{174}\Bigg)\\\\
 && \hspace{-0.8cm}{\psi_{171}}= 2\alpha \beta \zeta + \alpha^{2}(\zeta-1)Y_{2}+\beta^{2} \zeta Y_{2},\quad {\psi_{172}}=\alpha (2\beta_{2} \zeta R^{2}-4\beta_{1}(\zeta-4))+\beta Y_{2}(\beta_{2} \zeta R^{2}-2\beta_{1}(\zeta-5))\\\\
  && \hspace{-0.8cm}{\psi_{173}} = \beta ^2 Y_{2} \left(\beta_{2} \zeta (\zeta+2) R^2-2 \beta_{1}\left(\zeta^2-8 \zeta-1\right)\right)+2 \alpha  \beta  \zeta \left(\beta_{2} (\zeta+2) R^2-2 \beta_{1} (\zeta-7)\right)\nonumber\\&&\hspace{0.0cm} +\alpha ^2 (\zeta-1) Y_{2} \left(\beta_{2} \zeta R^2-2 \beta_{1}(\zeta-3)\right)\\\\
  && \hspace{-0.8cm}{\psi_{174}}=2 \alpha  \beta  \zeta \left(6 \beta_{1} \zeta+2 \beta_{2} \zeta R^2+\beta_{2} R^2\right)+2 \alpha ^2 (\zeta-1) Y_{2}\left(\beta_{1} (3 \zeta-1)+\beta_{2}\zeta R^2\right)\nonumber\\&&\hspace{0.0cm}+\beta ^2 \zeta Y_{2}\left(\beta_{1} (6 \zeta+2)+\beta_{2} (2 \zeta+1) R^2\right)\\\\
  && \hspace{-0.8cm}{\psi_{18}}=(\zeta+\xi R^{2})\beta \Bigg(\zeta \beta_{2} \psi_{181}+\xi^{3}R^{4} \beta^{2} \psi_{182}+\xi^{2} R^{2}\psi_{183}+\xi\psi_{184}\Bigg)\nonumber\\&&\hspace{0.0cm}+2(-1+\zeta)\alpha \Bigg(\zeta \beta_{2}\psi_{185} +\xi^{3} R^{3}\beta\psi_{186}+\xi^{2}R^{2}\psi{187}+\xi \psi_{188} \log\Bigg[\frac{\alpha}{\alpha+Y2\beta}\Bigg]\Bigg)\\\\
  && \hspace{-0.8cm}{\psi_{181}}=2 \alpha ^2 (\zeta-1)+\beta ^2 \zeta+3 \alpha  \beta  (\zeta-1) Y_{2},\quad {\psi_{182}}=\beta_{2} \zeta R^2-2 \beta_{1} (\zeta-7)\\\\
   && \hspace{-0.8cm}{\psi_{183}}=\beta ^2 \left(\beta_{1}\left(-2 \zeta^2+20 \zeta+6\right)+\beta_{2} \zeta (\zeta+2) R^2\right)-2 \alpha ^2 (\zeta-1) \left(2 \beta_{1} (\zeta-3)-\beta_{2} \zeta R^2\right)\nonumber\\&&\hspace{0.0cm}+\alpha  \beta  (\zeta-1)Y_{2} \left(\text{$\beta $1} (26-6 \zeta)+3 \beta_{2} \zeta R^2\right)\\\\
 \end{eqnarray*}
 \begin{eqnarray*}
    && \hspace{-0.8cm}{\psi_{184}}=4 \alpha ^2 (\zeta-1) \left(\beta_{1} (3\zeta-1)+\beta_{2} \zeta R^2\right)+\beta ^2 \zeta \left(6 \beta_{1} (\zeta+1)+\beta_{2} (2 \zeta+1) R^2\right)\nonumber\\&&\hspace{0.0cm}+2 \alpha  \beta  (\zeta-1) Y_{2}\left(\beta_{1}+9 \beta_{1} \zeta+3 \beta_{2}\zeta R^2\right)\\\\
    && \hspace{-0.8cm}{\psi_{185}}=2 \alpha  \beta  \zeta+\alpha ^2 (\zeta-1)Y_{2}+\beta ^2 \zeta Y_{2},\quad {\psi_{186}}=\alpha  \left(2 \beta_{2} \zeta R^2-4 \beta_{1} (\zeta-4)\right)+\beta  Y_{2} \left(\beta_{2} \zeta R^2-2 \beta_{1} (\zeta-5)\right)\\\\ 
   && \hspace{-0.8cm}{\psi_{187}}=\beta ^2 Y_{2} \left(\beta_{2} \zeta (\zeta+2) R^2-2 \beta_{1} \left(\zeta^2-8 \zeta-1\right)\right)+2 \alpha  \beta  \zeta \left(\beta_{2} (\zeta+2) R^2-2 \beta_{1} (\zeta-7)\right)\nonumber\\&&\hspace{0.0cm}+\alpha ^2 (\zeta-1)Y_{2} \left( \beta_{2} \zeta R^2-2 \beta_{1} (\zeta-3)\right)\\\\
    && \hspace{-0.8cm}{\psi_{188}}=2 \alpha  \beta  \zeta \left(6 \beta_{1} \zeta+2 \beta_{2}\zeta R^2+\beta_{2} R^2\right)\nonumber\\&&\hspace{0.0cm}+2 \alpha ^2 (\zeta-1)Y_{2} \left(\beta_{1}(3 \zeta-1+\beta_{2} \zeta R^2\right)+\beta ^2 \zeta Y_{2} \left(\beta_{1} (6 \zeta+2)+\beta_{2}(2 \zeta+1) R^2\right)\\\\
     &&\hspace{-0.8cm}e^{\upsilon}(r)=\Bigg((1-Y_{1}^{2})^{\frac{1}{4}}\Bigg(\frac{\alpha+\beta \times Y_{1}}{Y1}\Bigg(\frac{\alpha}{\beta^{3}}A_{1}\Bigg(\frac{\sec v \times \sec v -\cos v \times\cos v}{2}\Bigg)+\log\Bigg[cos v \times \cos v \Bigg]\Bigg) + B_{1} \Bigg) \Bigg)^{2}
\end{eqnarray*}

\end{document}